\documentclass[draftcls, onecolumn]{IEEEtran}
\usepackage{amsfonts}
\usepackage{amssymb}
\usepackage{graphicx}
\usepackage{amsmath}
\usepackage{bm}
\renewcommand{\baselinestretch}{1.0}
\begin{document}

\title{A semi-analytical model of Bilayer-Graphene Field Effect Transistor}
\author{Martina Cheli, Gianluca Fiori and Giuseppe Iannaccone\\
Dipartimento di Ingegneria dell'Informazione: Elettronica, Informatica, Telecomunicazioni, via Caruso 16, 56100 Pisa, Italy\\
email: {\tt \{martina.cheli, g.fiori, g.iannaccone\}@iet.unipi.it}}

\maketitle

\bibliographystyle{IEEEtran}



\begin{abstract}

Bilayer graphene has the very interesting property of an energy gap
tunable with the vertical electric field.
We propose an analytical model for a bilayer-graphene field-effect
transistor, suitable for exploring the
design parameter space in order to design a device structure with promising
performance in terms of transistor
operation. Our model, based on the effective mass approximation and
ballistic transport assumptions,
takes into account bilayer-graphene tunable gap and self-polarization,
and includes all
band-to-band tunneling current components, which are shown to
represent the major limitation to transistor operation,
because the  achievable energy gap is not sufficient to obtain
a large $I_{\rm on}/I_{\rm off}$ ratio.
\end{abstract}


\section{Introduction}
The progress of CMOS technology, with the pace foreseen by the 
International Technology Semiconductor Roadmap (ITRS)~\cite{ITRS2008}, 
cannot be based only on the capability to scale down device dimensions, 
but requires the introduction of new device architectures~\cite{Wang2007} 
and new materials for the channel, the gate stack and the contacts. This 
trend has already emerged for the recent technology nodes, and will hold 
-- probably requiring more aggressive innovations -- for devices
at the end of the Roadmap.
\\In the last decade carbon allotropes have attracted the attention of the scientific community, first 
with carbon nanotubes~\cite{Iijima91} and, since its isolation in 2004, with graphene~\cite{Novoselov2004}, which has  shown unique 
electronic~\cite{Neto2007} and physical properties~\cite{Geim2007}, such as unconventional integer quantum Hall effect~\cite{Novoselov2005, gusynin2005}, 
 high carrier mobility~\cite{Novoselov2004}  at room temperature,  
and  potential for a wide range of applications~\cite{Meyer2006,Scott2007,Dikin2007}, like nanoribbon FETs~\cite{Fiori2007}.
Despite graphene  is a  zero  gap material, an energy gap can   be   engineered 
 by "rolling" it in carbon nanotubes~\cite{Zhou99} or by the definition of lateral confinement like in graphene nanoribbons~\cite{Chen2007}. 
However, theoretical~\cite{Cohen2006} and experimental~\cite{Han2007} works have shown that
 significant gap in nanoribbons is obtained for widths close to 1-2 nm, which are  prohibitive for fabrication  technology on the scale of 
integrated circuits, at least in the medium term.
\\Recently, theoretical models~\cite{Nilsson2008,Castro2008,Mccann2006} and experiments~\cite{Ohta2006} have shown that bilayer graphene has 
the interesting property
of an energy gap tunable with an applied vertical electric field. Anyway, the largest attainable gap is of few hundreds of meV, which make 
its use questionable for 
nanoelectronics applications: limits and potentials  of bilayer graphene still have to be shown.  
\\From this point of view, device simulations can greatly help in 
assessing device performance. Bilayer-graphene
FETs (BG-FETs) have been compared against monolayer FETs, by means of the effective 
mass approximation~\cite{Ouyang2008} and Monte Carlo 
simulations~\cite{Harada2008} in the ballistic limit, showing really poor 
potential as compared to ITRS requirements~\cite{ITRS2008}. These 
approaches, however, did not take into account some of the main specific 
and important properties of bilayer graphene, such as the possibility of 
tuning the band gap and the dispersion relation with the vertical 
electric field, and dielectric polarization in the direction 
perpendicular to the 2D sheet. Such problems have been overcome in
Ref.~\cite{Fiori2008}, using a real space Tight-Binding approach.
However, for the limited set of device structures considered, the small 
band gap does not allow a proper on and off switching  of the transistor.

One limitation of detailed physical simulations
is that, despite their accuracy,  they  are typically  too demanding from a computational point of 
view for a complete investigation of device potential. Analytical approaches could  help in this 
case. One example has been proposed in Ref.~\cite{Ryzhii2008}, but it 
has serious drawbacks, because it completely neglects band-to-band 
tunneling and the dependence of the effective mass on the vertical 
electric field, providing a unrealistic optimistic picture of the 
achievable performance.
\\In this work, we have developed  a semi-analytical model for a  bilayer-graphene FET with two gates to study  the 
possibility of  realizing an FET by tuning the gap with a vertical electric field. 
The model has been validated through comparison with results obtained by means of a full 3D atomistic Poisson-Schr\"odinger
solver, showing good agreement in the applied bias range~\cite{Fiori2008, ViD}.
Interband tunneling proves to be the main limiting factor in device operation, as demonstrated by the device analysis performed in the parameter space.


\section{Model}
In this section we provide a detailed description of the developed model, which is based both on a top of the barrier model~\cite{natori} and on the calculation of all the interband tunneling components.
In particular we adopt the ballistic transport and the effective mass approximation, whose main electrical quantities, such as the effective mass and the energy gap, have been extracted from the energy bands obtained from a $p_z$-orbital Tight Binding (TB) Hamiltonian.
Since we want to address long channel devices, short channel effects have been completely neglected, as well as inelastic scattering mechanisms, which are expected to be negligible in this kind of material~\cite{Geim2007}.
With respect to more accurate atomistic models, the followed approach may underestimate the actual concentration of carriers in the channel, especially for large drain-to-source ($V_{DS}$) and gate voltages ($V_{GS}$), when parabolic band misses to match the exact dispersion relation. We however believe that the developed model represents a good trade-off between accuracy and speed. 
\subsection{ Effective mass approximation}
\begin{figure}[!h]
\vspace{1cm}
\centering
\includegraphics{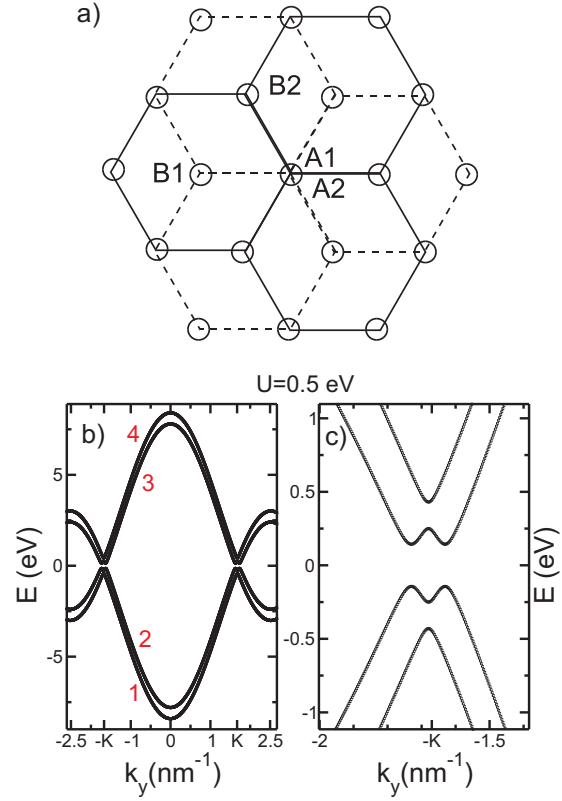}
\caption{a) Real space lattice structure of bilayer graphene. The bilayer consists of two coupled hexagonal lattices with inequivalent sites A1, B1 and A2, B2 in the first and in the second sheet, respectively, arranged according to Bernal (A2-B1) stacking.
b) Tight-Binding band structure of bilayer graphene for $U$=$U_1$-$U_2$=0.5~eV. c) Detail of the band structure in correspondence of band minimum {\bf{$k_{min}$}}: $K$ is the Dirac point.}
\label{fig1}
\end{figure}

In order to proceed with the definition of an analytical model based on the effective mass approximation, we first need an expression for the energy bands of  bilayer graphene. The top view of the bilayer-graphene lattice structure with carbon-carbon distance $a=1.44$~$\mathring{A}$ is shown in Fig.~\ref{fig1}(a): A1-B1 atoms lay on the top layer, while A2-B2 on the bottom layer. The energy dispersion relation can be computed by means of a  $p_{z}$-Tight Binding (TB) Hamiltonian~\cite{Wallace1947} considering two layers of graphene coupled in correspondence of the overlaying atoms A1 and A2.
\\The energy dispersion relation  reads~\cite{Nilsson2008} :
\begin{eqnarray}
E(\mathbf{k})=\frac{U_{1}+U_{2}}{2}\pm   
\end{eqnarray}
\begin{eqnarray} 
\sqrt{|f(\mathbf{k})|^2+ \frac{U^2}{4}+\frac{t_{\perp}^2}{2} \pm \frac{1}{2} \sqrt{4(U^2+t_{\perp}^2)|f(\mathbf{k})|^2 +t_{\perp}^4   }} \nonumber
\label{dispersion_relation}
,\end{eqnarray}
where $U_1$ and $U_2$ are the potential energies on the first and second layer, respectively, $U=U_1-U_2$, $t_{\perp}$=-0.35~eV is the inter-layer hopping parameter~\cite{Nilsson2008}, $\mathbf{k}=k_x\hat{k}_x+k_y\hat{k}_y $ and~\cite{Wallace1947}:
\begin{eqnarray}
f\mathbf(\mathbf{k})=te^{ik_{x}a/2}\left[2 \cos \left( \frac{k_{y}a\sqrt{3}}{2}\right)+e^{-i3k_{x}a/2}\right]   
\label{fdikappa}
,\end{eqnarray}
which is the well known off-diagonal element of the 2$\times$2  graphene $p_{z}$-Hamiltonian, where $t$ is the in-plane hopping parameter ($t$=-2.7~eV). 
In Fig.~\ref{fig1}(b) the band diagram for $U=0.5$~eV is shown. As can be seen, bilayer graphene has four bands, symmetric with respect to the coordinate axis. For large $U$, the ``mexican-hat'' behavior in correspondence of the band minima can be observed, as detailed in Fig.~\ref{fig1}(c).

Let us now consider the third band (Fig.~\ref{fig1}(b)), which corresponds to the conduction band (same considerations follow for the valence band, i.e. second band) and apply a parabolic band approximation in correspondence of the minimum $k_{min}$,  which reads~\cite{Nilsson2008}:
\begin{eqnarray}
k_{min}=\sqrt{\frac{U^2+2t_{\perp}^2}{U^2+t_{\perp}^2}}\frac{U}{2v_{F}\hbar}
\label{kminimo}
.\end{eqnarray}
The  dispersion relation can now be expressed as~\cite{Nilsson2008}
\begin{eqnarray}
E(\mathbf{k})=\frac{E_{gap}}{2}+\frac{\hbar^2}{2m^*}(|\mathbf{k}|-k_{min})^2+\frac{U_{1}+U_{2}}{2}
\label{Conduction_band_approx}
,\end{eqnarray}
where
\begin{eqnarray}
m^*=\frac{t_{\perp}(U^2+t_{\perp}^2)^{3/2}}{2U(U^2+2t_{\perp}^2)}\frac{1}{v_{F}^2} \quad \quad ; \quad \quad E_{gap}=\frac{Ut_{\perp}}{\sqrt{U^2+t_{\perp}^2}}
\label{massa}
,\end{eqnarray}
 $v_{F}=\frac{3at}{2\hbar}$ is the Fermi velocity and $\hbar$ is the reduced Planck's constant.

\begin{figure}[!t]
\vspace{1cm}
\centering
\includegraphics[scale=0.9]{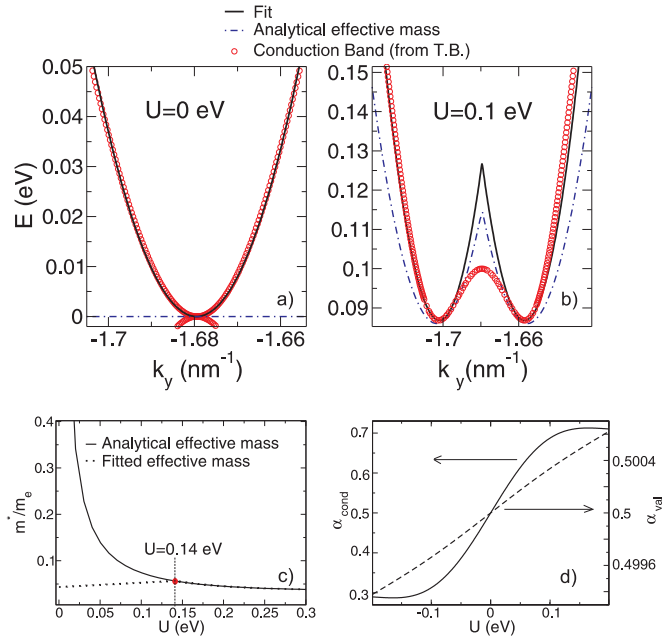}
\caption{Comparison  between energy dispersions obtained by means of the analytical effective mass (dashed-dotted line), fitted effective mass (solid line) and  TB Hamiltonian (circle)  for an  inter-layer  potential equal to  a) $U$=0~eV  and b) $U$=0.1~eV. c) Analytical  and  fitted relative effective mass as a function of the inter-layer potential $U$. $m_e$ is the free electron mass. d) $\alpha_{cond}(U)$ and  $\alpha_{val}(U)$  as a function of the inter-layer potential $U$. }
\label{fig2}
\end{figure}

As can be observed in  (\ref{massa}), the effective mass $m^*$ has a singularity for $U=0$, which is clearly unphysical.
In order to avoid such an issue, energy bands in the range $U \in [0, 0.14]$ have been fitted with the parabolic expression in (\ref{Conduction_band_approx}), within an energy range of 2$k_{B}$T from the band minimum (where $k_B$ is the Boltzmann constant and $T$ is the room temperature),  and using $m^*$ as a fitting parameter.
In  Figs.~\ref{fig2}(a), \ref{fig2}(b), we show, for two different inter-layer potential energies ($U$=0~eV and $U$=0.1~eV),  the TB energy bands as well as  the parabolic bands exploiting the analytical expression in (\ref{massa}) and the fitted values for $m^*$, respectively.
As can be seen,  the fitted effective mass manages to better match the TB band in the specified energy range.
In  Fig.~\ref{fig2}(c), we show the fitted effective mass for different $U$. In particular, for $U<0.14$~eV, $m^*$ can be expressed as:
\begin{eqnarray}
m^*=0.09U+0.043
\label{massa_fit}
,\end{eqnarray}
while for larger values eq.~(\ref{massa}) recovers.


\subsection{ Electrostatics}
Once obtained the expression for $m^*$, the electron concentration $n$ can be expressed as: 
\begin{eqnarray}
n=\frac{\nu}{2}\int_{E_{c}}^{+\infty} D(E) \left[f\left(E-E_{FS}\right)+f\left(E-E_{FD}\right)\right] dE 
\label{elettroni}
,\end{eqnarray}
where $f$ is the   Fermi-Dirac occupation factor, $E_{FS}$ and $E_{FD}$ are the Fermi energies of the source and drain, respectively, and $\nu$=2 is  band degeneracy.
$D(E)$ is the total density of states per unit area (for the complete calculation see the Appendix), which reads:
\begin{eqnarray}
 D(E)  =\frac{1}{2\pi\hbar}\left(  \frac{2m^*}{\hbar}+\sqrt{\frac{2m^*}{E-E_{c}}}k_{min}       \right),
\label{densitastati4}
\end{eqnarray}
where $E_{c}$ is the conduction band edge.
If we define:
\begin{eqnarray}
f_n(E_f)&=&\frac{m^*}{\pi\hbar^2}k_{B}T \ln \left[1+\exp\left(\frac{E_{c}-E_{f}}{k_{B}T} \right) \right]+ \nonumber \\
&&\frac{k_{min}
\sqrt{2m^*k_{B}T}}{2\pi\hbar} F_{1/2}\left(\frac{E_{c}- E_{f} }{k_{B}T}\right),
\label{elettroniintegrale}
\end{eqnarray}
where $F_{1/2}$  is the Fermi-Dirac integral of order $1/2$,
the electron concentration reads:
\begin{eqnarray}
n=\left[f_n(E_{FS})+f_n(E_{FD}) \right].
\label{elettroni2}
\end{eqnarray}
Analogous considerations can be made for the hole concentration $p$, which reads:
\begin{eqnarray}
p=\left[f_p(E_{FS})+f_p(E_{FD}) \right],
\label{lacune}
\end{eqnarray}
where
\begin{eqnarray}
f_p(E_f)&=&\frac{m^*}{\pi\hbar^2}k_{B}T \ln \left[1+\exp\left(\frac{E_{f}-E_{v}}{k_{B}T} \right) \right]\nonumber \\
&&+\frac{k_{min}
\sqrt{2m^*k_{B}T}}{2\pi\hbar} F_{1/2}\left(\frac{E_{f} -   E_{v}}{k_{B}T}\right),
\label{lacuneintegrale}
\end{eqnarray}
and $E_v$ is the valence band edge.
\\Once  $n$ and $p$ are computed,  attention has to be posed on how charge distributes on the two layers i.e. on dielectric polarization.
To this purpose,  we have  numerically extracted from TB simulations $\alpha_{val}(U)$ and  $\alpha_{cond}(U)$, that represent the fraction of the total states in the valence band and of electrons in the conduction band, respectively,  on layer 1~\cite{Fiori2008}. 
We computed $\alpha_{\rm cond}(U)$  for a particular bias ($U_1=-U_2=U/2$ and $E_F=0$~eV) and made the assumption that its dependence   on the bias can be neglected. As far as $\alpha_{\rm val}$ is concerned, we assumed in our considered bias range, that all electron states in the valence band are fully occupied and therefore $f(E)=1$.
Fig.~\ref{fig2}(d), shows $\alpha_{cond}(U)$ and  $\alpha_{val}(U)$  as a function of the inter-layer potential $U$. The charge density $\rho_j$ per unit area on  layer $j$ ($j$=1,2) is expressed as the sum of the polarization charge, electrons and holes and finally reads:
 \begin{eqnarray}\label{rho}
 \rho_{1}(U)&=& q\left\{\left[1-2\alpha_{\rm val}(U)\right]N_{tot} \right. \nonumber \\ 
 && \left. -n\left[1-2\alpha_{\rm cond}(U)\right]+p\alpha_{\rm cond}(U) \right\} ;\\
 \rho_{2} (U)&=&q\left\{\left[2\alpha_{\rm val}(U)-1\right]N_{tot} \right.\nonumber \\
 && \left.-n\alpha_{\rm cond}(U)+p\left[1-2\alpha_{\rm cond}(U)\right]\right\} ,\nonumber
\end{eqnarray}
where $q$ is the electron charge and $N_{tot}$ is the concentration of ions per unit area.

\begin{figure}[!h]
\vspace{1cm}
\centering
\includegraphics[scale=0.4]{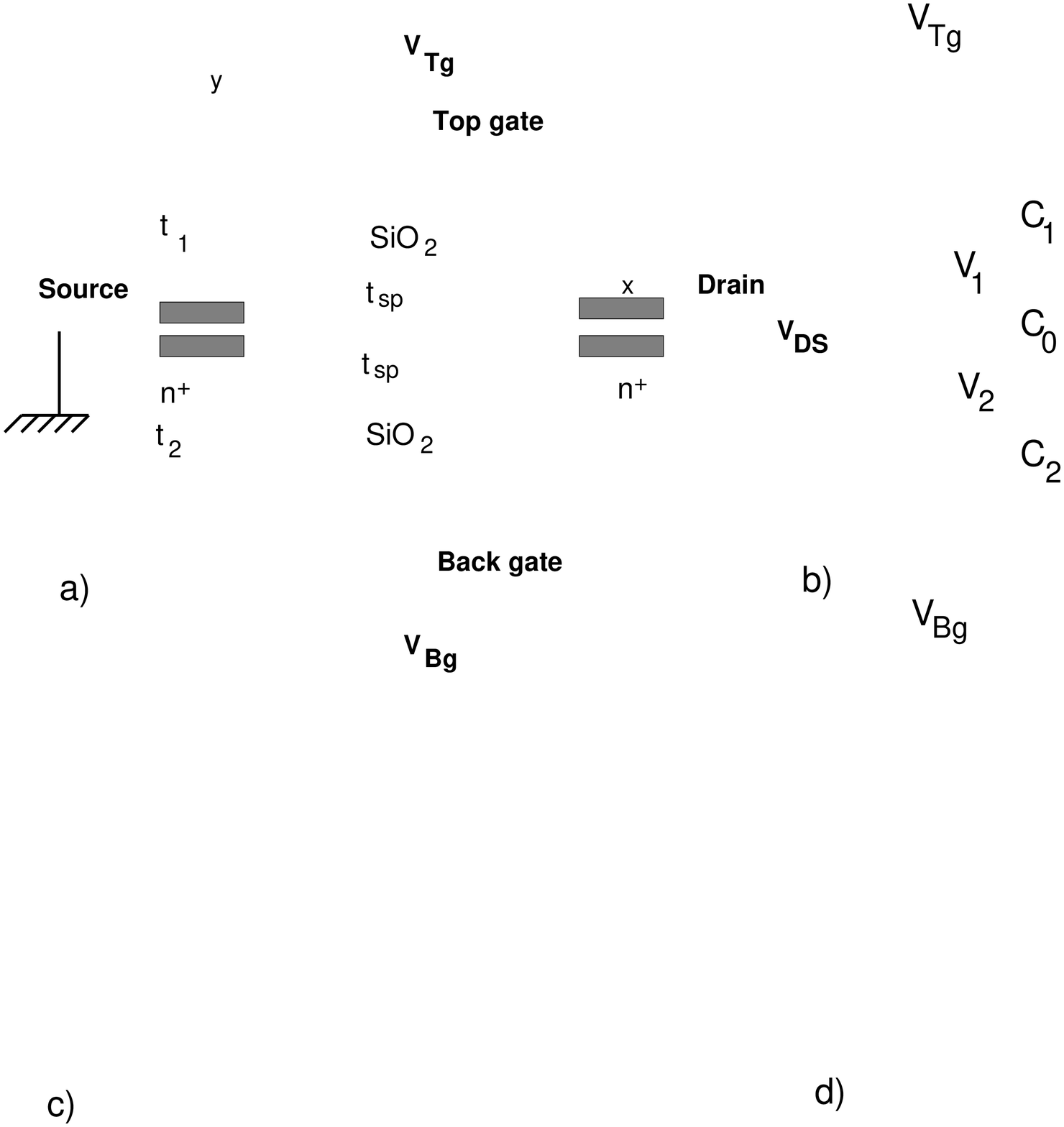}
\caption{a) Sketch of the considered  bilayer-graphene FET: $d$ is the inter-layer distance,  $t_{1}$ and $t_{2}$ are the top and back oxide thicknesses. b) Equivalent capacitance circuit of the simulated device. c) Flat band diagram of the BG-FET along the $y$ direction. d) Conduction and valence band edge profiles in the longitudinal direction; we assume that deep in the source and drain regions, the electric field induced by the gate vanishes and the gap gradually reduces to zero. }
\label{fig3}
\normalsize
\end{figure}

The considered device structure  is a double-gate FET embedded in SiO$_2$.  The bilayer graphene  inter-layer distance  $d$ is equal to 0.35 nm, while two different  oxide  thicknesses  $t_{1}$ and $t_{2}$ 
have been considered (Fig.~\ref{fig3}(a)). An air interface between bilayer graphene and oxide has also been taken into account  ($t_{sp}=$0.5~nm)~\cite{Dai2008}.
For such a system, we can define an equivalent capacitance circuit as in Fig.~\ref{fig3}(b), where $C_{0}$=$\frac{\epsilon_{0}}{d}$, 
$C_{1}$=$\Big[ \frac{t_{1}}{\epsilon_{1}}+\frac{t_{sp}}{\epsilon_{0}}\Big]^{-1}$, 
$C_{2}$=$\Big[ \frac{t_{2}}{\epsilon_{2}}+\frac{t_{sp}}{\epsilon_{0}}\Big]^{-1}$ and  $\epsilon_{1}$= $\epsilon_{2}$= 3.9$\epsilon_{0}$, while $\epsilon_{0}=8.85~\times10^{-12}$~F/m. $V_{Tg}$ and $V_{Bg}$ are the top gate and back gate voltage respectively, $V_{1} \equiv \frac{-U_1}{q}$ and $V_{2} \equiv \frac{-U_2}{q}$.
In  Fig.~\ref{fig3}(c), the flat band diagram along the transverse direction ($y$ axis) is shown. Metal work functions for the back gate and top gate are equal to 4.1~eV [$\Phi_{Bg}$=$\Phi_{Tg}$=4.1~eV], while the graphene work function ($\Phi_{gra}$) is equal to 4.5~eV~\cite{Wildoer1998}.  
 E$_{FTg}$, E$_{FBg}$ are the Fermi level of the top and of the back gate, respectively.

The conduction band edge inserted in eq.~(\ref{elettroniintegrale}), can be expressed as: 

\begin{eqnarray}
E_{c}=\Phi_{Bg}+E_{FBg}-\Phi_{gra}+\frac{U_{1}+U_{2}}{2}+\frac{E_{gap}}{2}
\label{Conduction_band_channel}
.\end{eqnarray}

Applying the Gauss theorem, we obtain the following expression:
\begin{equation} 
 \left\{
 \begin{array}{rl}
  C_{1}  (V_{Tg}-V_{1})+   (V_{2}-V_{1})C_{0}=-\rho_{1}   \\
  C_{0}  (V_{1}-V_{2})+  (V_{Bg} -V_{2}) C_{2}=-\rho_{2}. 
 \end{array} \right.
 \label{self consistent}
\end{equation}
Eqs. (\ref{rho}) and (\ref{self consistent}) are then solved self-consistently till convergence on $V_1$ and $V_2$ is achieved.

\subsection{ Current}
Drain-to-source ($J_{TOT}$) current is computed at the end of the self-consistent scheme.
As depicted in Fig.~\ref{fig3}(d), $J_{TOT}$ 
 consists of three different components: the first  is due to the  thermionic current $J_{th}$ over the barrier~\cite{natori}, whereas the second ($J_{TS}$) and the third ($J_{TD}$) to
band-to-band tunneling. 
In the same picture, we sketch the conduction band edge $E_{CS}$ ($E_{CD}$) and the valence band edge $E_{VS}$ ($E_{VD}$) at the source (drain).
Assuming reflectionless contacts, the thermionic current is due to electrons injected from the source with positive velocity $v_{x}>0$   and to electrons injected from the drain with $v_{x}<0$:
 \begin{eqnarray}
J_{th}&=&\frac{-q}{\pi^2\hbar} \int_{-\infty}^{+\infty} dk_{y}\left[\int_{k_{x}^{>}}\frac{\partial E}{\partial k_{x}} f(E-E_{FS})dk_{x}+ \right. \nonumber \\
&& \left. \int_{k_{x}^{<}}\frac{\partial E}{\partial k_{x}} f(E-E_{FD})   dk_{x}   \right]
\label{corrente_totale_integrale1}
,\end{eqnarray}
where $E=E_c+\frac{\hbar^2}{2m^*}\left(|\mathbf{k}|-k_{min}  \right)^2$,  $v_x= \frac{1}{\hbar}\frac{\partial E}{\partial k_{x}}$ is the group velocity and  $k_{x}^>$ ($k_{x}^<$) is the wavevector range for which $v_{x}>0$ ($v_{x}<0$). For the complete derivation see the Appendix.

Let us now discuss the band-to-band tunneling current due to the  barrier at source(drain) contact, which reads:
\begin{eqnarray}
J_{Ti}&=&2\int_{k_{y}}\int_{k_{x}^{>}} q \frac{1}{2\pi^2} \frac{1}{\hbar}\frac{\partial E}{\partial k_{x}} T_i(k_{y}) \left[f(E-E_{FS}) \right.\nonumber \\
&& \left. -f(E-E_{FD})\right] \partial k_{x} \partial k_{y} \qquad i=S,D,
\label{tunnelingint}
\end{eqnarray}
where $S$ refers to the source and $D$ to the drain, while  $T_i(k_{y})$ is the transmission coefficient at the different reservoirs. The key issue in computing  (\ref{tunnelingint}) is the definition of an expression for $T_i(k_{y})$, which accounts for  band-to-band tunneling process.

We have assumed a  non charge-neutrality region of fixed width $\Delta x$ at the contact/channel interface  and an electric field  
 $\mathcal{E}_i$=$(E_{c}-E_{Fi})/(q\Delta x )$ with $i=S,D$. For what concern the $J_{TS}$ term,  electrons emitted with electrochemical potential $E_{FS}$ see two triangular barriers, one at the source junction and one in correspondence of the drain 
(Fig.~\ref{fig3}(d)), whose heights are equal to $E_{gap}$ and width $W_i=E_{gap}/(q\mathcal{E}_i)$. 
Assuming the same $\Delta x$ for both source and drain junctions,  the drain barrier is transparent with respect to the source barrier, since, for large $V_{DS}$, the electric field at the source  is smaller than the electric field at the drain barrier:  $T_S(k_y)$ is therefore essentially given by the source junction barrier. Same considerations follow for the other band-to-band tunneling current component $J_{TD}$, flowing only  through  the drain-channel contact. In this case $\mathcal{E}_D=E_c-E_{FD}=\frac{E_{c}-E_{FS}+qV_{DS}}{\Delta x q}$.

Assuming the WKB approximation, the transmission coefficient can be expressed as:
\begin{eqnarray}
T_i(k_{y})=e^{-2 \int_{[W_{b}]}{|Im\{k_x\}| dx}} , \qquad i=S,D
\label{transmission}
,\end{eqnarray}
where $Im\{k_{x}\}$ is the imaginary part of $k_x$ and  is obtained from:
\begin{eqnarray}
\frac{\hbar^2}{2m^*}\left(|\mathbf{k}|-k_{min} \right)^2=q\mathcal{E}_ix-E_{gap} , \qquad i=S,D.
\label{TK}
\end{eqnarray}

Finally, $J_{Ti}$ is computed performing the integral (\ref{tunnelingint}) numerically.

\section{ Exploration of the design space}
In order to validate our model, we have first compared analytical results with those obtained by means of numerical NEGF Tight Binding simulations~\cite{ViD}, considering a test structure
with $t_{1}=t_{2}=$1.5~nm, $t_{sp}$= 0.5~nm, $\Phi_{gra}=\Phi_{G}=\Phi_{Bg}$=4.1~eV, $V_{DS}$=0.1~V and $V_{Bg}$=0~V.
\begin{figure}[!h]
\vspace{1cm}
\centering
\includegraphics{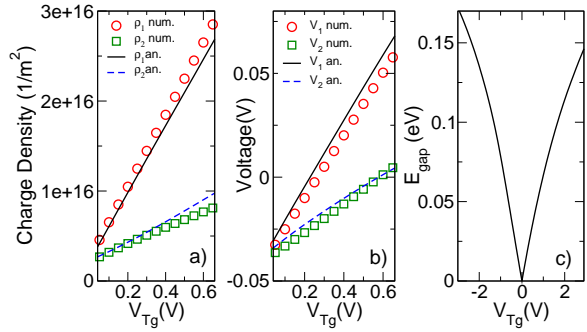}
\caption{Comparison between analytical and numerical simulation of a) $\rho_1$ and $\rho_2$ and b) $V_1$ and $V_2$ as a function of $V_{Tg}$. $V_{Bg}$=0~V and $V_{DS}$=0~V. c) Energy gap as a function of top gate voltage, with $V_{Bg}=0$~V.}
\label{fig4}
\normalsize
\end{figure}
In Fig.~\ref{fig4}(a)-(b)  the electron concentrations ($\rho_1$, $\rho_2$) and the electrostatic potentials ($V_1$, $V_2$) on layer 1 and 2 are shown, as a function of $V_{Tg}$, for $V_{DS}$=0~V and $V_{Bg}$=0~V.    
As can be seen, results are in good agreement. Some discrepancies however occur for larger $V_{DS}$ ($V_{DS}>0.2~V$), where the parabolic band approximation misses to reproduce band behavior for large $k_{y}$. 
In Fig.~\ref{fig4}(c)  the energy gap is plotted as a function of $V_{Tg}$. As can be seen, even for large $V_{Tg}$, the biggest attainable $E_{gap}$ is close to 0.15~eV.

\begin{figure}[!h]
\vspace{1cm}
\centering
\includegraphics{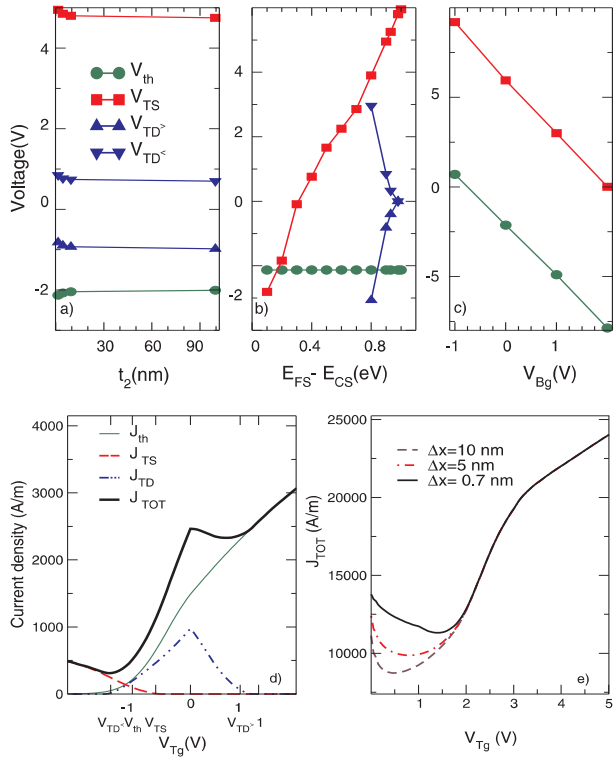}
\caption{a) Thresholds as a function of the oxide thickness $t_{2}$ for $V_{DS}$= 0.5 V, $V_{BG}$=0 V, $E_{FS}-E_{CS}$=0.9 eV. b) Thresholds as a  function of $E_{FS}-E_{CS}$ for $t_{2}$=1.5 nm,  $V_{DS}$= 0.5 V, $V_{BG}$=0 V. c) Thresholds as a function of  $V_{Bg}$ for $t_{2}$=1.5 nm,  $V_{DS}$= 0.5 V and  $E_{FS}-E_{CS}$= 1 eV.  
d) Total current $J_{TOT}$ and its three components ($J_{TS}$, $J_{TD}$, $J_{th}$) for  $V_{DS}$=0.1 V, $V_{Bg}$=0 V and $E_{FS}-E_{CS}$= 0.5 eV . Thresholds are shown along the coordinate axis. e) Total current for $t_{2}$=1.5 nm,  $V_{DS}$= 0.5 V, $E_{FS}-E_{CS}$= 1 eV and $V_{BG}$=0 V and different $\Delta x$. }
\label{fig5}
\normalsize
\end{figure}

Let us now consider  the different contributions of the three current components ($J_{th}$, $J_{TS}$ and  $J_{TD}$) to the total current $J_{TOT}$ (Fig.~\ref{fig5}(d)). For each of these components, we can define a sort of threshold voltage, above which their contribution is not negligible. In particular, $J_{th}$ starts to be relevant as soon as $E_{c}\sim E_{FS}$. We then define  $V_{th}$ as the $V_{Tg}$ for which $E_{c}=E_{FS}$. Similarly, interband current $J_{TS}$ is not zero when $E_v$$\ge$$E_{CS}$ so we define $V_{TS}$ the top-gate voltage for which $E_v$$=$$E_{CS}$. Finally, $J_{TD}$ is not zero in the energy range $E_{CD}< E < E_{VS}$: we  define $V_{TD}^>$ and $V_{TD}^<$ the top-gate voltages for which  $E_{VS}$$=$$E_{CD}$; thanks to these definitions, we can qualitatively evaluate current contribution by observing   the band structure. 

Our goal is indeed to obtain the largest value for the $I_{\rm on}/I_{\rm off}$ ratio, and this is only possible if the band-to-band component of the current is suppressed. We have considered three different solutions to accomplish this task: by varying the back gate oxide ($t_{2}$), by varying the $E_{FS}-E_{CS}$ or $E_{FD}-E_{CD}$ difference, or by simply varying the back gate voltage. If otherwise specified, $\Delta x=0.7$~nm, as obtained from TB simulations of an abrupt junction with the same doping of the considered BG-FET. In Fig.~\ref{fig5}(a)-(b)-(c)   the above-defined thresholds are shown for the three considered cases. 
\\As shown in  Fig.~\ref{fig5}(a), back gate oxide thickness has no effect in our case, since the top layer screens the electric field induced by the top gate, as can also be seen from Fig.~\ref{fig4}(b), where $V_2$ remains almost constant. Fig.~\ref{fig5}(b) shows thresholds as a function of ($E_{FS}-E_{CS}$), and therefore as a function of  dopant concentration.
We observe that for $E_{FS}-E_{CS}=1$~eV, $V_{TD}^>=V_{TD}^<$, so that $J_{TD}$ is practically eliminated, while $J_{TS}$ increases since $V_{TS}$ becomes larger.
In Fig.~\ref{fig5}(c), we show $V_{th}$ and $V_{TS}$, for $E_{FS}-E_{CS}=1~eV$, as a function of $V_{Bg}$. Unfortunately the two curves have the same behavior, so that $V_{TS}$ cannot be reduced to values smaller than $V_{th}$, or --in other words--  we cannot  
suppress current due to interband tunneling at source contact.

\begin{figure}[!h]
\vspace{1cm}
\centering
\includegraphics[scale=0.4]{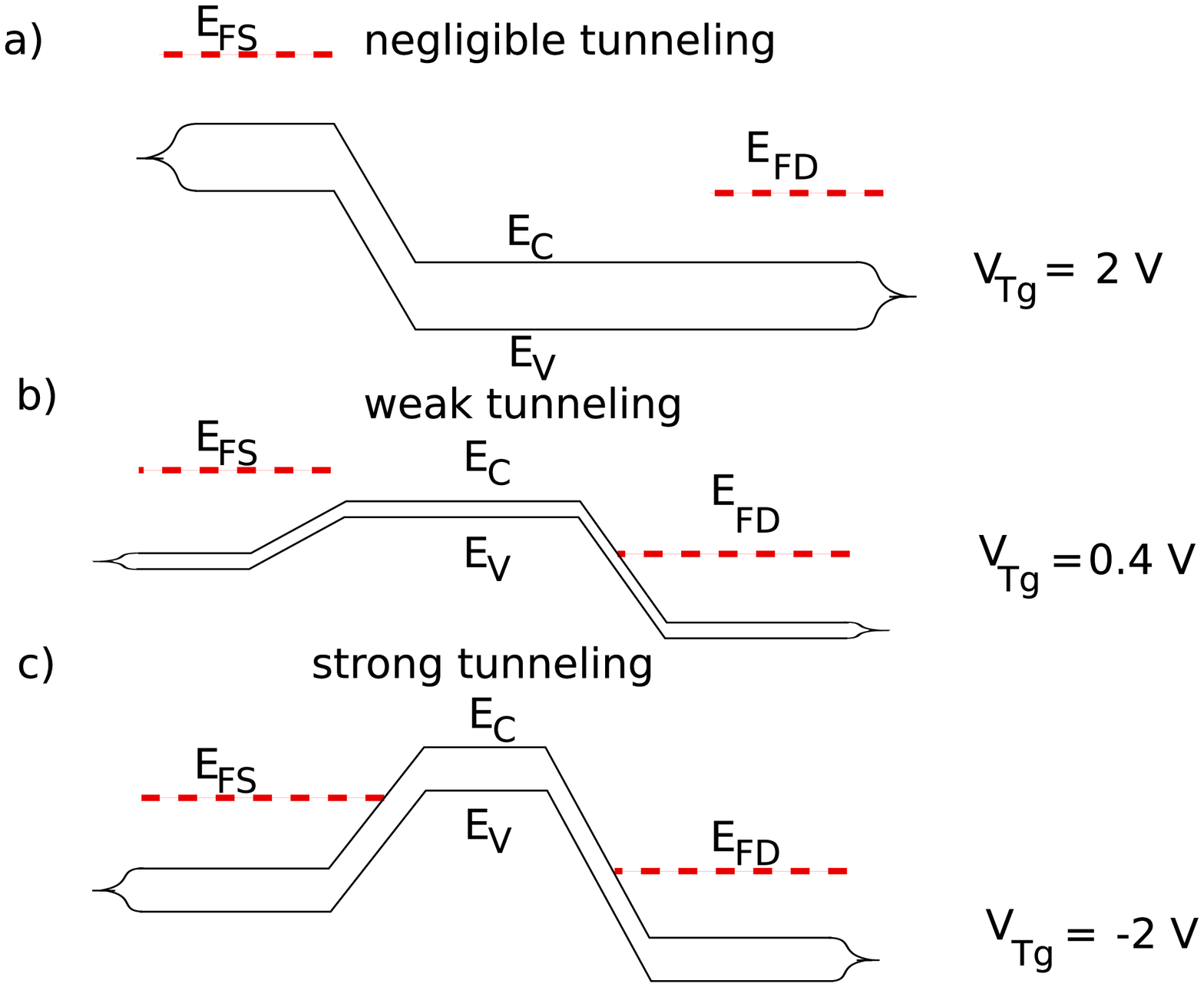}
\caption{Simulated band edges when: a) Tunnel current weakly affects $J_{TOT}$; b) $J_{TS}+J_{TD}$ is one order of magnitude smaller than the  total current; c) Inter band tunneling current is  the dominant  component.}
\label{figNUOVA}
\normalsize
\end{figure}

We have then computed the transfer characteristics  for $V_{Bg}=0 V$, $t_{1}=t_{2}=$1.5 nm, $E_{FS}-E_{CS}$=1 eV. In Fig.~\ref{fig5}(e)   $J_{TOT}$ is shown.  As can be seen, poor   $I_{\rm on}/I_{\rm off}$ ratio  can be obtained since  band-to-band tunneling at source contact is too high  as also observed in  graphene FET~\cite{Ryzhii2008bis}. 
Reducing $\mathcal{E}$, i.e. $T(k_y)$, could lead to a reduction of $J_{TS}$  and consequently to   an improvement of the $I_{\rm on}/I_{\rm off}$ ratio.  As can be seen in Fig.~\ref{fig5}(e),  an improved  $I_{\rm on}/I_{\rm off}$ is obtained increasing  $\Delta x$ to  5-10 nm,
but it is still lower than the ITRS requirements ($10^4$) for digital circuits.
In Fig.~\ref{figNUOVA}, we also sketch the simulated band edges for three different cases: a) when tunneling is negligible (
$J_{TOT}\simeq J_{Ti}$ for $V_{Tg}=2$~V), b) when tunneling weakly affect the total current ($J_{TOT}\simeq 10 (J_{TS}+J_{TD})$ for 
$V_{Tg}=0.4$~V) and c) when tunneling represents the predominant component ($J_{TOT}\simeq (J_{TS}+J_{TD})$ for $V_{Tg}=-2$~V).

\section{ Conclusion}
We have developed an analytical model for bilayer-graphene field
effect transistors, suitable for the exploration of the design
parameter space. The model is based on some simplifying assumptions,
such as the effective mass approximation, but includes all the
relevant physics of bilayer graphene. First and foremost, it
includes the tunable gap of bilayer graphene with the vertical
electric field, which is exploited in order to induce the largest gap,
when the device is in the off state. It also fully includes
polarization of bilayer graphene in response to a vertical electric
field. As far as transport is concerned, it includes the thermionic
current components and all interband tunneling components, which are
the main limiting factor in achieving a large $I_{\rm on}/I_{\rm
off}$ ratio. Significant aspects of the model have been validated
through comparisons with numerical TB NEGF simulations.

Due to the small computational requirements, we have been able to
explore the parameter design space of  bilayer-graphene FETs in order
to maximize the $I_{\rm on}/I_{\rm off}$ ratio. Despite applied
vertical field manages to induce an energy gap of the order of one
hundred meV, band-to-band tunneling greatly affects device performance,
limiting its use for device applications. A larger gap must be
induced to make bilayer graphene a useful channel material for 
digital applications, probably by combining different options, such as using
bilayer graphene in addition to limited lateral confinement, stress,
or doping.

\appendices
\section{}

\subsection{Density of states}
The total density of states  can be computed as follow, performing the integral over the first Brillouin zone (BZ):
\begin{equation}
 D(E)  = \frac{2}{(2\pi)^2}\int_{BZ}{ \delta\left(E_{c}+\frac{\hbar^2}{2m^*}(|\mathbf{k}|-k_{min})^2-E\right)2\pi kdk}.
\label{densitastati1}
\end{equation}
\\If we apply the following property of the delta function
\begin{eqnarray}
\delta[f(x)]=\sum_{n} \frac{\delta(x-x_{n})}{|f'(x_{n})|},
\end{eqnarray}
where $x_{n}$ are the  zeroes of the function $f(x)$,  eq. (\ref{densitastati1}) reads:
\begin{eqnarray}
 D(E)&=& \frac{1}{\pi}\frac{\sqrt{2m^*}}{\hbar} \frac{1}{2\sqrt{E-E_{c}}}\int_{BZ'}{ \delta\left(r-\sqrt{E-E_{c}}\right) }\nonumber \\
 && {\left(\frac{\sqrt{2m^*}}{\hbar}r+k_{min}\right) dr} \nonumber \\
 	&&=\frac{1}{2\pi\hbar}\left(  \frac{2m^*}{\hbar}+\sqrt{\frac{2m^*}{E-E_{c}}}k_{min}       \right),
\label{densitastati4}
\end{eqnarray}
where  r=$\frac{\hbar}{\sqrt{2m^*}}(|\mathbf{k}|$-$k_{min})$.

\subsection{Thermionic Current}
In order to derive the expression of the thermionic current, we have first to compute the group velocity $v_x$, which reads:
\begin{eqnarray}
\frac{1}{\hbar}\frac{\partial E(k_{x}, k_{y})}{\partial k_{x}} &=&\frac{\mathbf{1}}{\hbar}\frac{\partial}{\partial k_{x}} \left( \frac{E_{gap}}{2}+ \frac{\hbar^2}{2m^*} \left(|\mathbf{k}|-k_{min} \right)^2 \right) \nonumber \\
&& = \frac{\hbar k_{x}}{m^{*}}\left( 1-\frac{k_{min}}{|\mathbf{k}|}  \right)	
\label{derivata}
.\end{eqnarray}
Replacing (\ref{derivata}) in (\ref{corrente_totale_integrale1}) we obtain:
\begin{eqnarray}
J_{th}&=&\frac{-q\hbar}{\pi^2 m^*}\int_{-\infty}^{+\infty}dk_{y}\left[\int_{k_{x}^{>}}k_{x}\left(1-\frac{k_{min}}{|\mathbf{k}|}\right)f(E-E_{FS})dk_{x} \right.\nonumber \\
&& \left. +\int_{k_{x}^{<}}k_{x}\left(1-\frac{k_{min}}{|\mathbf{k}|}\right)f(E-E_{FD})dk_{x}\right]
\label{corrente_totale_integrale2}
.\end{eqnarray}
In order to remove the singularity in eq. (\ref{corrente_totale_integrale2}) for $k_x=k_y=0$,  we can use  cylindrical coordinates, i.e. $k_{x}$=$k$$cos\theta$,  $k_{y}$=$k$$sin\theta$ and $|\mathbf{k}|$=$k$. In this representation, the condition $v_{x}>0$ translates in:
\begin{equation*}
 \left\{
 \begin{array}{rl}
  k cos\theta>0\\
  k-k_{min}>0
\end{array} \right. \qquad
\left\{
\begin{array}{rl}
k cos\theta<0\\
k-k_{min}<0
\end{array} \right.
\label{estremivxpositiva}
\end{equation*}
The integral (\ref{corrente_totale_integrale2}) becomes:
\begin{align}
J_{th} &=\frac{2q\hbar}{\pi^2 m^*} \int_{0}^{k_{max}}k \left(k-k_{min}\right) \left[f_{S}(E,k)-f_{D}(E, k) \right]dk.
\label{corrente_totale_integrale_polare}
\end{align}

\subsection{Transmission coefficient}
The tunneling transmission probability $T(k_{y})$ has been computed through the WKB approximation. The $|Im\{k_x\}|$ in eq.~(\ref{transmission})
 is computed from the energy  dispersion relation as follows: from (\ref{TK}) we can write
\begin{eqnarray}
\left(\sqrt{k_{x}^2+k_{y}^2}-k_{min} \right)^2=\frac{2m^*}{\hbar^2}(q\mathcal{E}x-\frac{E_{gap}}{2}).
\label{tun1}
\end{eqnarray}
Defining
\begin{eqnarray}
\beta(x)=\frac{\sqrt{2m^*\left( \frac{E_{gap}}{2}-q\mathcal{E}x\right)}}{\hbar}>0
\label{beta}
\end{eqnarray}
and inserting eq.~(\ref{beta}) in  eq.~(\ref{tun1}), we obtain:
\begin{eqnarray}
\left(k_{x}^2+k_{y}^2\right)^{\frac{1}{2}}-k_{min}=i\beta(x),
\label{tun2}
\end{eqnarray}
which  reads:
\begin{eqnarray}
k_{x}^2=\left(k_{min}^2-\beta(x)^2-k_{y}^2 \right)+2i\beta(x)k_{min}.
\label{tun3}
\end{eqnarray}
If we expressed $k_x$ as $k_{x}$=$a+ib$ ($a,b \in \Re$), $k_x^2$ reads:
\begin{eqnarray}
k_{x}^2=a^2-b^2+2iab.
\label{tun4}
\end{eqnarray}
By comparing eq.~(\ref{tun3}) and eq.~(\ref{tun4}), $|Im\{k_x\}|$ simply reads:
\begin{eqnarray}
|Im\{k_x\}|=\sqrt{\frac{C+\sqrt{C^2+4\beta(x)^2k_{min}^2}}{2}}
\label{C}
,\end{eqnarray}
with $C=-k_{min}^2+\beta^2(x)+k_{y}^2$.

{\bf Acknowledgements}
The work was supported in part by the EC Seventh Framework Program under project GRAND (Contract
215752) and  by the Network of Excellence NANOSIL (Contract 216171).
Authors gratefully acknowledge Network for Computational Nanotechnology (NCN) for providing computational resources at nanohub.org, through which part of the results here shown has been obtained.

\bibliography{BibliographyArticleNano}

\begin{thebibliography}{10}
\providecommand{\url}[1]{#1}
\csname url@rmstyle\endcsname
\providecommand{\newblock}{\relax}
\providecommand{\bibinfo}[2]{#2}
\providecommand\BIBentrySTDinterwordspacing{\spaceskip=0pt\relax}
\providecommand\BIBentryALTinterwordstretchfactor{4}
\providecommand\BIBentryALTinterwordspacing{\spaceskip=\fontdimen2\font plus
\BIBentryALTinterwordstretchfactor\fontdimen3\font minus
  \fontdimen4\font\relax}
\providecommand\BIBforeignlanguage[2]{{%
\expandafter\ifx\csname l@#1\endcsname\relax
\typeout{** WARNING: IEEEtran.bst: No hyphenation pattern has been}%
\typeout{** loaded for the language `#1'. Using the pattern for}%
\typeout{** the default language instead.}%
\else
\language=\csname l@#1\endcsname
\fi
#2}}

\bibitem{ITRS2008}
\BIBentryALTinterwordspacing
``International {T}echnology {R}oadmap for {S}emiconductor 2007.'' [Online].
  Available: \url{http://public.itrs.net.}
\BIBentrySTDinterwordspacing

\bibitem{Wang2007}
Z.~F. Wang, H.~Zheng, Q.~W. Shi, and J.~Chen, ``Emerging nanocircuit paradigm:
  Graphene-based electronics for nanoscale computing,'' in \emph{Nanoscale
  Architectures, 2007. NANOARCH 2007. IEEE International Symposium on}, Oct.
  2007, pp. 93--100.

\bibitem{Iijima91}
S.~Iijima, ``Helical microtubules of graphitic carbon,'' \emph{Nature}, vol.
  354, no. 6348, pp. 56--58, Nov. 1991.

\bibitem{Novoselov2004}
K.~S. Novoselov, A.~K. Geim, S.~V. Morozov, D.~Jiang, Y.~Zhang, S.~V. Dubonos,
  I.~V. Grigorieva, and A.~A. Firsov, ``Electric field effect in atomically
  thin carbon films,'' \emph{Science}, vol. 306, no. 5696, pp. 666--669, Oct.
  2004.

\bibitem{Neto2007}
A.~H.~C. Neto, F.~Guinea, N.~M.~R. Peres, K.~S. Novoselov, and A.~K. Geim,
  ``The electronic properties of graphene,'' \emph{Rev. Mod. Phys.}, vol.~81,
  pp. 109--163, Jan. 2009.

\bibitem{Geim2007}
A.~K. Geim and K.~S. Novoselov, ``The rise of graphene,'' \emph{Nat Mater},
  vol.~6, no.~3, pp. 183--191, Mar. 2007.

\bibitem{Novoselov2005}
K.~S. Novoselov, A.~K. Geim, S.~V. Morozov, D.~Jiang, M.~I. Katsnelson, I.~V.
  Grigorieva, S.~V. Dubonos, and A.~A. Firsov, ``Two-dimensional gas of
  massless {D}irac fermions in graphene,'' \emph{Nature}, vol. 438, no. 7065,
  pp. 197--200, Nov. 2005.

\bibitem{gusynin2005}
V.~P. Gusynin and S.~G. Sharapov, ``{U}nconventional {I}nteger {Q}uantum {H}all
  effect in graphene,'' \emph{Phys. Rev. Lett.}, vol.~95, no.~14, pp.
  146\,801--146\,805, Sep. 2005.

\bibitem{Meyer2006}
J.~C. Meyer, A.~K. Geim, M.~I. Katsnelson, K.~S. Novoselov, T.~J. Booth, and
  S.~Roth, ``The structure of suspended graphene sheets,'' \emph{Nature}, vol.
  446, no. 7131, pp. 60--63, Dec. 2006.

\bibitem{Scott2007}
J.~S. Bunch, A.~M. Van~der Zande, S.~S. Verbridge, I.~W. Frank, D.~M.
  Tanenbaum, J.~M. Parpia, H.~G. Craighead, , and P.~L. McEuen,
  ``Electromechanical {R}esonators from {G}raphene {S}heets,'' \emph{Science},
  vol. 315, no. 5811, pp. 490--493, Jan. 2007.

\bibitem{Dikin2007}
D.~A. Dikin, S.~Stankovich, E.~J. Zimney, R.~D. Piner, G.~H.~B. Dommett,
  G.~Evmenenko, S.~T. Nguyen, and R.~S. Ruoff, ``Preparation and
  characterization of graphene oxide paper,'' \emph{Nature}, vol. 448, pp.
  457--460, Jul. 2007.

\bibitem{Fiori2007}
G.~Fiori and G.~Iannaccone, ``Simulation of {G}raphene {N}anoribbon
  {F}ield-{E}ffect {T}ransistors,'' \emph{IEEE Electron Device Letters},
  vol.~28, no.~8, pp. 760--762, Aug. 2007.

\bibitem{Zhou99}
C.~Zhou, J.~Kong, and H.~Dai, ``Electrical measurements of individual
  semiconducting single-walled nanotubes of various diameters,'' \emph{Appl.
  Phys. Lett.}, vol.~76, no.~12, pp. 1597--1599, Mar. 2000.

\bibitem{Chen2007}
Z.~Chen, Y.-M. Lin, M.~J. Rooks, and P.~Avouris, ``Graphene nano-ribbon
  electronics,'' \emph{Phys. E}, vol.~40, no.~2, pp. 228 -- 232, Jun. 2007.

\bibitem{Cohen2006}
Y.-W. Son, M.~L. Cohen, and S.~G. Louie, ``Energy gaps in graphene
  nanoribbons,'' \emph{Phys. Rev. Lett.}, vol.~97, no.~21, pp.
  216\,803--216\,806, Nov. 2006.

\bibitem{Han2007}
M.~Y. Han, B.~\"{O}zyilmaz, Y.~Zhang, and P.~Kim, ``Energy {B}and-{G}ap
  {E}ngineering of {G}raphene {N}anoribbons,'' \emph{Phys. Rev. Lett.},
  vol.~98, no.~20, pp. 206\,805--206\,809, May 2007.

\bibitem{Nilsson2008}
J.~Nilsson, C.~A.~H. Neto, F.~Guinea, and N.~M.~R. Peres, ``Electronic
  properties of bilayer and multilayer graphene,'' \emph{Phys. Rev. B},
  vol.~78, no.~4, pp. 045\,405--1--34, Jul. 2008.

\bibitem{Castro2008}
E.~V. Castro, K.~S. Novoselov, S.~V. Morozov, N.~M.~R. Peres, J.~M.~B.~L. dos
  Santos, J.~Nilsson, F.~Guinea, A.~K. Geim, and A.~H.~C. Neto, ``Biased
  bilayer graphene: semiconductor with a gap tunable by the electric field
  effect,'' \emph{Phys. Rev. Lett.}, vol.~99, no.~21, p. 216802, Nov. 2007.

\bibitem{Mccann2006}
E.~McCann, ``Asymmetry gap in the electronic band structure of bilayer
  graphene,'' \emph{Phys. Rev. B}, vol.~74, no.~16, pp. 161\,403--161\,407,
  Nov. 2006.

\bibitem{Ohta2006}
T.~Ohta, A.~Bostwick, T.~Seyller, K.~Horn, and E.~Rotenberg, ``Controlling the
  {E}lectronic {S}tructure of {B}ilayer {G}raphene,'' \emph{Science}, vol. 313,
  no. 5789, pp. 951--954, Aug. 2006.

\bibitem{Ouyang2008}
Y.~Ouyang, P.~Campbell, and J.~Guo, ``Analysis of ballistic monolayer and
  bilayer graphene field-effect transistors,'' \emph{Appl. Phys. Lett.},
  vol.~92, no.~92, pp. 063\,120--063\,122, Feb. 2008.

\bibitem{Harada2008}
N.~Harada, M.~Ohfuti, and Y.~Awano, ``Performance estimation of {G}raphene
  {F}ield-{E}ffect-{T}ransistors using semiclassical {M}onte {C}arlo
  {S}imulation,'' \emph{Appl. Phys. Exp.}, vol.~1, no.~1, pp.
  024\,002--024\,004, Feb. 2008.

\bibitem{Fiori2008}
G.~Fiori and G.~Iannaccone, ``On the possibility of tunable-gap bilayer
  graphene {FET},'' \emph{IEEE Electron Device Letters}, vol.~30, no.~3, pp.
  261--264, Mar. 2009.

\bibitem{Ryzhii2008}
V.~Ryzhii, M.~Ryzhii, A.~Satou, T.~Otsuji, and N.~Kirova, ``Device model for
  graphene bilayer field-effect transistor,'' \emph{J. App. Phys.}, vol. 105,
  no.~10, p. 104510, Dec. 2009.

\bibitem{ViD}
\BIBentryALTinterwordspacing
``Nanotcad vides.'' [Online]. Available: \url{Code and {D}ocumentation can be
  found at the url: http://www.nanohub.org/tools/}
\BIBentrySTDinterwordspacing

\bibitem{natori}
K.~Natori, ``Ballistic metal-oxide-semiconductor field effect transistor,''
  \emph{J. App. Phys.}, vol.~76, no.~8, pp. 4879--4890, Jul. 1994.

\bibitem{Wallace1947}
P.~R. Wallace, ``The band theory of graphite,'' \emph{Phys. Rev.}, vol.~71,
  no.~9, pp. 622--634, May 1947.

\bibitem{Dai2008}
X.~Li, X.~Wang, L.~Zhang, S.~Lee, and H.~Dai, ``{Chemically Derived,
  Ultrasmooth Graphene Nanoribbon Semiconductors},'' \emph{Science}, vol. 319,
  no. 1229, p. 1150878, Feb. 2008.

\bibitem{Wildoer1998}
J.~W.~G. Wildoer, L.~C. Venema, A.~G. Rinzler, R.~E. Smalley, and C.~Dekker,
  ``Electronic structure of atomically resolved carbon nanotubes,''
  \emph{Nature}, vol. 391, pp. 59--62, Jan. 1998.

\bibitem{Ryzhii2008bis}
V.~Ryzhii, M.~Ryzhii, and T.~Otsuji, ``Thermionic and tunneling transport
  mechanism in graphene field-effect transistors,'' \emph{Phys. Stat. Sol.
  (a)}, vol. 205, no.~7, pp. 1527--1533, Mar. 2008.

\end{thebibliography}


@article{Iijima91,
	author = {Iijima, S.},
	journal = {Nature},
	month = {Nov.},
	number = {6348},
	pages = {56--58},
	posted-at = {2007-04-23 16:57:35},
	priority = {0},
	title = {Helical microtubules of graphitic carbon},
	volume = {354},
	year = {1991}
}



@article{Nilsson2008,
	author = {Nilsson, Johan   and Neto, Castro  A. H.  and Guinea, F.  and Peres, N. M. R.},
	journal = {Phys. Rev. B},
	number = {4},
	pages={045405-1--34},
	month={Jul.},
	title = {Electronic properties of bilayer and multilayer graphene},
	volume = {78},
	year = {2008}
}

@article{Geim2007,
	author = {Geim, A. K.  and Novoselov, K. S. },
	citeulike-article-id = {1154381},
	journal = {Nat Mater},
	keywords = {graphene},
	month = {Mar.},
	number = {3},
	pages = {183--191},
	posted-at = {2007-03-11 21:07:03},
	priority = {3},
	title = {The rise of graphene},
	volume = {6},
	year = {2007}
}

@ARTICLE{Neto2007,
  author = {A.~H.~Castro Neto and F. Guinea and N.~M.~R. Peres and K.~S. Novoselov and A.~K. Geim},
  title = {The electronic properties of graphene},
  journal = {Rev.  Mod. Phys.},
  month={Jan.},
  volume = {81},
  pages = {109--163},
  year = {2009}
}


@article{Meyer2006,
	author = {Meyer, Jannik  C.  and Geim, A. K.  and Katsnelson, M. I.  and Novoselov, K. S.  and Booth, T. J.  and Roth, S. },
	citeulike-article-id = {1134992},
	eprint = {cond-mat/0701379},
	issn = {0028-0836},
	journal = {Nature},
	keywords = {graphene},
	number = {7131},
	month={Dec.},
	pages = {60--63},
	posted-at = {2007-09-10 09:09:08},
	priority = {3},
	publisher = {Nature Publishing Group},
	title = {The structure of suspended graphene sheets},
	volume = {446},
	year={2006}
}

@article{Castro2008,
  author = {Eduardo V. Castro and K.~S. Novoselov and S.~V. Morozov and N.~M.~R. Peres and J.~M.~B.~Lopes dos Santos and Johan Nilsson and F. Guinea and A.~K. Geim and A.~H.~Castro Neto},
  title = {Biased  bilayer graphene: semiconductor with a gap tunable by the electric field effect},
  journal={Phys. Rev. Lett.},
  volume = {99},
  number = {21},
  pages={216802},
  month={Nov.},
  year = {2007}
}




@article{Geim2004,
	author = {Novoselov, K. S.  and Geim, A. K.  and Morozov, S. V.  and Jiang, D.  and Zhang, Y.  and Dubonos, S. V.  and Grigorieva, I. V.  and Firsov, A. A. },
	citeulike-article-id = {1507653},
	journal = {Science},
	keywords = {graphene},
	month = {Oct.},
	number = {5696},
	pages = {666--669},
	posted-at = {2007-07-27 22:45:03},
	priority = {2},
	title = {Electric {F}ield {E}ffect in {A}tomically {T}hin {C}arbon {F}ilms},
	volume = {306},
	year = {2004}
}

	

@article{Novoselov2005,
	author = {Novoselov, K. S.  and Geim, A. K.  and Morozov, S. V.  and Jiang, D.  and Katsnelson, M. I.  and Grigorieva, I. V.  and Dubonos, S. V.  and Firsov, A. A. },
	citeulike-article-id = {386216},
	issn = {0028-0836},
	journal = {Nature},
	keywords = {graphene},
	month={Nov.},
	number = {7065},
	pages = {197--200},
	posted-at = {2006-09-14 18:49:12},
	priority = {0},
	publisher = {Nature Publishing Group},
	title = {Two-dimensional gas of massless {D}irac fermions in graphene},
	volume = {438},
	year = {2005}
}

@article{Jiang2007,
	author = {Jiang, Z.  and Zhang, Y.  and Tan, Y. W.  and Stormer, H. L.  and Kim, P. },
	booktitle = {Exploring graphene - Recent research advances},
	citeulike-article-id = {2176722},
	journal = {Solid State Communications},
	keywords = {2d, nano, quantum-Hall-effect},
	month = {July},
	number = {1-2},
	pages = {14--19},
	posted-at = {2007-12-28 07:52:37},
	priority = {2},
	title = {Quantum Hall effect in graphene},
	volume = {143},
	year = {2007}
}

@article{Berger2006,
	author = {Berger, Claire   and Song, Zhimin   and Li, Xuebin   and Wu, Xiaosong   and Brown, Nate   and Naud, Cecile   and Mayou, Didier   and Li, Tianbo   and Hass, Joanna   and Marchenkov, Alexei  N.  and Conrad, Edward  H.  and First, Phillip  N.  and de Heer, Walt  A. },
	citeulike-article-id = {679510},
	journal = {Science},
	keywords = {graphene},
	month = {May},
	number = {5777},
	pages = {1191--1196},
	posted-at = {2006-07-13 19:17:29},
	priority = {2},
	title = {Electronic Confinement and Coherence in Patterned Epitaxial Graphene},
	volume = {312},
	year = {2006}
}

@article{Fiori2007,
	author = {Fiori, G.  and Iannaccone, G. },
	booktitle = {Electron Device Letters, IEEE},
	citeulike-article-id = {3007895},
	doi = {http://dx.doi.org/10.1109/LED.2007.901680},
	journal = {IEEE Electron Device Letters},
	keywords = {diplomarbeit, from\_norbert, gnrs, graphene, theory, tight-binding},
	number = {8},
	month={Aug.},
	pages = {760--762},
	posted-at = {2008-07-16 10:10:02},
	priority = {3},
	title = {Simulation of {G}raphene {N}anoribbon {F}ield-{E}ffect {T}ransistors},
	volume = {28},
	year = {2007}
}

		

@article{Wang2008,
	author = {Wang, Xinran   and Ouyang, Yijian   and Li, Xiaolin   and Wang, Hailiang   and Guo, Jing   and Dai, Hongjie  },
	citeulike-article-id = {2817533},
	journal = {Phys. Rev. Lett.},
	keywords = {experimental, gnr, graphene, transport},
	number = {20},
	posted-at = {2008-05-20 20:51:33},
	priority = {5},
	publisher = {APS},
	title = {Room-Temperature All-Semiconducting Sub-10-nm Graphene Nanoribbon Field-Effect Transistors},
	volume = {100},
	year = {2008}
}

@inproceedings{Wang2007,
	author = {Wang, Z. F.  and Zheng, Huaixiu   and Shi, Q. W.  and Chen, Jie  },
	booktitle = {Nanoscale Architectures, 2007. NANOARCH 2007. IEEE International Symposium on},
	title = {Emerging nanocircuit paradigm: Graphene-based electronics for nanoscale computing},
	citeulike-article-id = {2548117},
	journal = {Nanoscale Architectures, 2007. NANOARCH 2007. IEEE International Symposium on},
	keywords = {graphene},
	month={Oct.},
	pages = {93--100},
	posted-at = {2008-03-18 03:15:20},
	priority = {2},
	year = {2007}
}
	

@article{Liang2008,
	author = {Liang, G.   and Neophytos, N.   and Lundstrom, M.S.   and Nikonov, D.E.  },
	journal = {Nano Lett.},
	number = {7},
	pages = {1819--1824},
	title = {Contact Effects in Graphene Nanoribbon Transistors},
	volume = {8},
	year = {2008}
}


@article{Ohta2006,
	author = {Ohta, Taisuke   and Bostwick, Aaron   and Seyller, Thomas   and Horn, Karsten   and Rotenberg, Eli  },
	citeulike-article-id = {1372203},
	journal = {Science},
	keywords = {graphene},
	month = {Aug.},
	number = {5789},
	pages = {951--954},
	posted-at = {2007-06-08 10:26:26},
	priority = {2},
	title = {Controlling the {E}lectronic {S}tructure of {B}ilayer {G}raphene},
	volume = {313},
	year = {2006}
}

	
@article{Mccann2006,
	author = {McCann, Edward  },
	citeulike-article-id = {915490},
	journal = {Phys. Rev. B},
	keywords = {bandstructure, bilayer, graphene},
	number = {16},
	month={Nov.},
	pages = {161403--161407},
	posted-at = {2006-10-27 19:50:47},
	priority = {2},
	publisher = {APS},
	title = {Asymmetry gap in the electronic band structure of bilayer graphene},
	volume = {74},
	year = {2006}
}

@Article{Wallace1947,
  title = {The Band Theory of Graphite},
  author = {Wallace, P. R.},
  journal = {Phys. Rev.},
  volume = {71},
  number = {9},
  pages = {622--634},
  numpages = {12},
  year = {1947},
  month = {May},
  publisher = {American Physical Society}
}
	
@article{Wildoer1998,
	abstract = {Not Available},
	author = {Wildoer, J. W. G.  and Venema, L. C.  and Rinzler, A. G.  and Smalley, R. E.  and Dekker, C. },
	citeulike-article-id = {1810243},
	journal = {Nature},
	keywords = {nanotubes, of, singel, transport, wall},
	month = {Jan.},
	pages = {59--62},
	posted-at = {2007-10-23 11:45:18},
	priority = {2},
	title = {Electronic structure of atomically resolved carbon nanotubes},
	volume = {391},
	year = {1998}
}

	

@article{Ouyang2008,
	author ={Ouyang, Y.  and Campbell, P.  and Guo, J. },
	journal = {Appl. Phys. Lett.},
	number = {92},
	pages = {063120--063122},
	month={Feb.},
	title = {Analysis of ballistic monolayer and bilayer graphene field-effect transistors},
	volume = {92},
	year = {2008}
}

@article{Harada2008,
	author ={Harada, N.  and Ohfuti, M.  and Awano, Y. },
	journal = {Appl. Phys. Exp.},
	number = {1},
	pages = {024002--024004},
	month={Feb.},
	title = {Performance estimation of {G}raphene {F}ield-{E}ffect-{T}ransistors using semiclassical {M}onte {C}arlo {S}imulation},
	volume = {1},
	year = {2008}
}	


@article{Fiori2008,
  author = {Fiori, G. and  Iannaccone, G.},
  title = {On the possibility of tunable-gap bilayer graphene {FET}},
  journal= {IEEE Electron Device Letters},
  month={Mar.},
  number={3},
  pages = {261--264},
  volume = {30},
  year = {2009}
}


@article{Scott2007,
	author={ Bunch,J. Scott and  Van der Zande, Arend M. and   Verbridge, Scott S.  and Frank, Ian W.    and  Tanenbaum,  David M. and   Parpia, Jeevak M. and  Craighead, Harold G. and  and  McEuen, Paul L.},
	journal={Science},
	number={5811},
	month={Jan.},
	pages={490--493},
	title={Electromechanical {R}esonators from {G}raphene {S}heets},
	volume={315},
	year={2007}
	}
	
	
	
	
@article{Dikin2007,
	author = {Dikin, Dmitriy  A.  and Stankovich, Sasha   and Zimney, Eric  J.  and Piner, Richard  D.  and Dommett, Geoffrey  H. B.  and Evmenenko, Guennadi   and Nguyen, Sonbinh  T.  and Ruoff, Rodney  S. },
	journal = {Nature},
	pages = {457--460},
	month={Jul.},
	title = {Preparation and characterization of graphene oxide paper},
	volume = {448},
	year={2007}
}
	

@article{Chen2007,
title = "Graphene nano-ribbon electronics",
journal = "Phys. E",
volume = "40",
number = "2",
pages = "228 - 232",
year = "2007",
month= "Jun.",
issn = "1386-9477",
author = "Zhihong Chen and Yu-Ming Lin and Michael J. Rooks and Phaedon Avouris",
keywords = "Graphene",
keywords = "FET",
keywords = "Semiconducting",
keywords = "Edge states"
}

@article{Zhou99,
	author = {Zhou, C. and  Kong, J. and Dai, H. },
	journal = {Appl. Phys. Lett.},
	pages = {1597--1599},
	month={Mar.},
	number={12},
	title = {Electrical measurements of individual semiconducting single-walled nanotubes of various diameters},
	volume = {76},
	year={2000}
}

 

@article{Cohen2006,
author = {Young-Woo Son and Marvin L. Cohen and Steven G. Louie},
collaboration = {},
title = {Energy Gaps in Graphene Nanoribbons},
publisher = {APS},
year = {2006},
journal = {Phys. Rev. Lett.},
volume = {97},
month={Nov.},
number = {21},
eid = {216803},
numpages = {4},
pages = {216803--216806},
}

 


@article{Han2007,
author = {Melinda Y. Han and Barbaros \"{O}zyilmaz and Yuanbo Zhang and Philip Kim},
collaboration = {},
title = {Energy {B}and-{G}ap {E}ngineering of {G}raphene {N}anoribbons},
publisher = {APS},
year = {2007},
journal = {Phys. Rev. Lett.},
volume = {98},
month={May},
number = {20},
eid = {206805},
numpages = {4},
pages = {206805--206809},
}

@ARTICLE{Rhew2002,
   author = {{Rhew}, J. and {Ren}, Z. and {Lundstrom}, M.},
    title = "{A numerical study of ballistic transport in a nanoscale MOSFET}",
  journal = {Solid State Electronics},
     year = {2002},
    month = nov,
   volume = {46},
    pages = {1899-1906},
      doi = {10.1016/S0038-1101(02)00130-2},
   adsurl = {http://adsabs.harvard.edu/abs/2002SSEle..46.1899R},
  adsnote = {Provided by the SAO/NASA Astrophysics Data System}
}

@misc{ViD,
title={NanoTCAD ViDES},
url={Code and {D}ocumentation can be found at the url: http://www.nanohub.org/tools/ }
}

 
@ARTICLE{gusynin2005,
  author = {V.~P. Gusynin and S.~G. Sharapov},
  title = {{U}nconventional {I}nteger {Q}uantum {H}all effect in graphene},
  journal = {Phys. Rev. Lett.},
  volume = {95},
  month={Sep.},
  number={14},
  pages = {146801--146805},
  year = {2005}
}
 
@article{Novoselov2004,
	abstract = {We describe monocrystalline graphitic films, which are a few atoms thick but are nonetheless stable under ambient conditions, metallic, and of remarkably high quality. The films are found to be a two-dimensional semimetal with a tiny overlap between valence and conductance bands, and they exhibit a strong ambipolar electric field effect such that electrons and holes in concentrations up to 1013 per square centimeter and with room-temperature mobilities of [\~{}]10,000 square centimeters per volt-second can be induced by applying gate voltage. 10.1126/science.1102896},
	author = {Novoselov, K. S.  and Geim, A. K.  and Morozov, S. V.  and Jiang, D.  and Zhang, Y.  and Dubonos, S. V.  and Grigorieva, I. V.  and Firsov, A. A. },
	citeulike-article-id = {1507653},
	journal = {Science},
	keywords = {electronic-properties, experiment, graphene, science},
	month = {Oct.},
	number = {5696},
	pages = {666--669},
	posted-at = {2008-07-07 19:44:25},
	priority = {2},
	title = {Electric Field Effect in Atomically Thin Carbon Films},
	volume = {306},
	year = {2004}
}

	 
 @article{natori,
author = {Kenji Natori},
collaboration = {},
title = {Ballistic metal-oxide-semiconductor field effect transistor},
publisher = {AIP},
year = {1994},
journal = {J. App. Phys.},
volume = {76},
month={Jul.},
number = {8},
pages = {4879--4890},
keywords = {MOSFET; TRANSPORT PROCESSES; BALLISTICS; IV CHARACTERISTIC; CARRIER DENSITY; OPTICAL PHONONS; SILICON},
doi = {10.1063/1.357263}
}


@misc{ITRS2008,
title={International {T}echnology {R}oadmap for {S}emiconductor 2007},
url={ http://public.itrs.net.}
}
 
 
@article{Dai2008,
author = {Li, Xiaolin and Wang, Xinran and Zhang, Li and Lee, Sangwon and Dai, Hongjie},
title = {{Chemically Derived, Ultrasmooth Graphene Nanoribbon Semiconductors}},
journal = {Science},
volume = {319},
month={Feb.},
number = {1229},
pages = {1150878},
doi = {10.1126/science.1150878},
year = {2008}

} 
 
 
@article{Ryzhii2008,
author = {V. Ryzhii and M. Ryzhii and A. Satou and T. Otsuji and N. Kirova},
collaboration = {},
title = {Device model for graphene bilayer field-effect transistor},
publisher = {AIP},
year = {2009},
journal = {J. App. Phys.},
month={Dec.},
volume = {105},
number = {10},
eid = {104510},
numpages = {9},
pages = {104510},
keywords = {field effect transistors; graphene; molecular electronics},
}

@article{Ryzhii2008bis,
	author = {Ryzhii, Victor   and Ryzhii, Maxim  and Otsuji, Taiichi},
	journal = {Phys. Stat. Sol. (a)},
	number = {7},
	pages={1527--1533},
	month={Mar.},
	title = {Thermionic and tunneling transport mechanism in graphene field-effect transistors},
	volume = {205},
	year = {2008}
}


%
%
%
%
%
%
%
%
%
%
%
%
%
%
%
%
%
%
%
%
%
%
%






@STRING{IEEE_J_AES        = "{IEEE} Trans. Aerosp. Electron. Syst."}
@STRING{IEEE_J_ANE        = "{IEEE} Trans. Aerosp. Navig. Electron."}
@STRING{IEEE_J_ANNE       = "{IEEE} Trans. Aeronaut. Navig. Electron."}
@STRING{IEEE_J_AS         = "{IEEE} Trans. Aerosp."}
@STRING{IEEE_J_AIRE       = "{IEEE} Trans. Airborne Electron."}
@STRING{IEEE_J_MIL        = "{IEEE} Trans. Mil. Electron."}



@STRING{IEEE_J_ITS        = "{IEEE} Trans. Intell. Transport. Syst."}
@STRING{IEEE_J_VT         = "{IEEE} Trans. Veh. Technol."}
@STRING{IEEE_J_VC         = "{IEEE} Trans. Veh. Commun."}



@STRING{IEEE_J_SPL        = "{IEEE} Signal Processing Lett."}
@STRING{IEEE_J_ASSP       = "{IEEE} Trans. Acoust., Speech, Signal Processing"}
@STRING{IEEE_J_AU         = "{IEEE} Trans. Audio"}
@STRING{IEEE_J_AUEA       = "{IEEE} Trans. Audio Electroacoust."}
@STRING{IEEE_J_AC         = "{IEEE} Trans. Automat. Contr."}
@STRING{IEEE_J_CAS        = "{IEEE} Trans. Circuits Syst."}
@STRING{IEEE_J_CASVT      = "{IEEE} Trans. Circuits Syst. Video Technol."}
@STRING{IEEE_J_CASI       = "{IEEE} Trans. Circuits Syst. {I}"}
@STRING{IEEE_J_CASII      = "{IEEE} Trans. Circuits Syst. {II}"}
@STRING{IEEE_J_CT         = "{IEEE} Trans. Circuit Theory"}
@STRING{IEEE_J_CST        = "{IEEE} Trans. Contr. Syst. Technol."}
@STRING{IEEE_J_SP         = "{IEEE} Trans. Signal Processing"}
@STRING{IEEE_J_SU         = "{IEEE} Trans. Sonics Ultrason."}
@STRING{IEEE_J_SAP        = "{IEEE} Trans. Speech Audio Processing"}
@STRING{IEEE_J_UE         = "{IEEE} Trans. Ultrason. Eng."}
@STRING{IEEE_J_UFFC       = "{IEEE} Trans. Ultrason., Ferroelect., Freq. Contr."}



@STRING{IEEE_J_COML       = "{IEEE} Commun. Lett."}
@STRING{IEEE_J_JSAC       = "{IEEE} J. Select. Areas Commun."}
@STRING{IEEE_J_COM        = "{IEEE} Trans. Commun."}
@STRING{IEEE_J_COMT       = "{IEEE} Trans. Commun. Technol."}
@STRING{IEEE_J_WCOM       = "{IEEE} Trans. Wireless Commun."}



@STRING{IEEE_J_ADVP       = "{IEEE} Trans. Adv. Packag."}
@STRING{IEEE_J_CHMT       = "{IEEE} Trans. Comp., Hybrids, Manufact. Technol."}
@STRING{IEEE_J_CPMTA      = "{IEEE} Trans. Comp., Packag., Manufact. Technol. {A}"}
@STRING{IEEE_J_CPMTB      = "{IEEE} Trans. Comp., Packag., Manufact. Technol. {B}"}
@STRING{IEEE_J_CPMTC      = "{IEEE} Trans. Comp., Packag., Manufact. Technol. {C}"}
@STRING{IEEE_J_CAPT       = "{IEEE} Trans. Comp. Packag. Technol."}
@STRING{IEEE_J_CAPTS      = "{IEEE} Trans. Comp. Packag. Technol."}
@STRING{IEEE_J_CPART      = "{IEEE} Trans. Comp. Parts"}
@STRING{IEEE_J_EPM        = "{IEEE} Trans. Electron. Packag. Manufact."}
@STRING{IEEE_J_MFT        = "{IEEE} Trans. Manufact. Technol."}
@STRING{IEEE_J_PHP        = "{IEEE} Trans. Parts, Hybrids, Packag."}
@STRING{IEEE_J_PMP        = "{IEEE} Trans. Parts, Mater., Packag."}



@STRING{IEEE_J_TCAD       = "{IEEE} J. Technol. Computer Aided Design"}
@STRING{IEEE_J_CAD        = "{IEEE} Trans. Computer-Aided Design"}



@STRING{IEEE_J_IT         = "{IEEE} Trans. Inform. Theory"}
@STRING{IEEE_J_KDE        = "{IEEE} Trans. Knowledge Data Eng."}



@STRING{IEEE_J_C          = "{IEEE} Trans. Comput."}
@STRING{IEEE_J_ECOMP      = "{IEEE} Trans. Electron. Comput."}
@STRING{IEEE_J_EVC        = "{IEEE} Trans. Evol. Comput."}
@STRING{IEEE_J_FUZZ       = "{IEEE} Trans. Fuzzy Syst."}
STRING{IEEE_J_MC          = "{IEEE} Trans. Mobile Comput."}
@STRING{IEEE_J_NET        = "{IEEE/ACM} Trans. Networking"}
@STRING{IEEE_J_NN         = "{IEEE} Trans. Neural Networks"}
@STRING{IEEE_J_PDS        = "{IEEE} Trans. Parallel Distrib. Syst."}
@STRING{IEEE_J_SE         = "{IEEE} Trans. Software Eng."}



@STRING{IEEE_J_IP         = "{IEEE} Trans. Image Processing"}
@STRING{IEEE_J_MM         = "{IEEE} Trans. Multimedia"}
@STRING{IEEE_J_VCG        = "{IEEE} Trans. Visual. Comput. Graphics"}



@STRING{IEEE_J_JRA        = "{IEEE} J. Robot. Automat."}
@STRING{IEEE_J_HFE        = "{IEEE} Trans. Hum. Factors Electron."}
@STRING{IEEE_J_MMS        = "{IEEE} Trans. Man-Mach. Syst."}
@STRING{IEEE_J_PAMI       = "{IEEE} Trans. Pattern Anal. Machine Intell."}
@STRING{IEEE_J_RA         = "{IEEE} Trans. Robot. Automat."}
@STRING{IEEE_J_SMC        = "{IEEE} Trans. Syst., Man, Cybern."}
@STRING{IEEE_J_SMCA       = "{IEEE} Trans. Syst., Man, Cybern. {A}"}
@STRING{IEEE_J_SMCB       = "{IEEE} Trans. Syst., Man, Cybern. {B}"}
@STRING{IEEE_J_SMCC       = "{IEEE} Trans. Syst., Man, Cybern. {C}"}
@STRING{IEEE_J_SSC        = "{IEEE} Trans. Syst. Sci. Cybernetics"}



@STRING{IEEE_J_GE         = "{IEEE} Trans. Geosci. Electron."}
@STRING{IEEE_J_GRS        = "{IEEE} Trans. Geosci. Remote Sensing"}
@STRING{IEEE_J_OE         = "{IEEE} J. Oceanic Eng."}



STRING{IEEE_J_CJECE       = "Canadian J. Elect. Comput. Eng."}
@STRING{IEEE_J_PROC       = "Proc. {IEEE}"}
@STRING{IEEE_J_EDU        = "{IEEE} Trans. Educ."}
@STRING{IEEE_J_EM         = "{IEEE} Trans. Eng. Manage."}
STRING{IEEE_J_EWS         = "{IEEE} Trans. Eng. Writing Speech"}
@STRING{IEEE_J_PC         = "{IEEE} Trans. Prof. Commun."}



@STRING{IEEE_J_AWPL       = "{IEEE} Antennas Wireless Propagat. Lett."}
@STRING{IEEE_J_MGWL       = "{IEEE} Microwave Guided Wave Lett."}
@STRING{IEEE_J_MWCL       = "{IEEE} Microwave Wireless Compon. Lett."}
@STRING{IEEE_J_AP         = "{IEEE} Trans. Antennas Propagat."}
@STRING{IEEE_J_EMC        = "{IEEE} Trans. Electromagn. Compat."}
@STRING{IEEE_J_MAG        = "{IEEE} Trans. Magn."}
@STRING{IEEE_J_MTT        = "{IEEE} Trans. Microwave Theory Tech."}
@STRING{IEEE_J_RFI        = "{IEEE} Trans. Radio Freq. Interference"}
@STRING{IEEE_J_TJMJ       = "{IEEE} Transl. J. Magn. Jpn."}



@STRING{IEEE_J_EC         = "{IEEE} Trans. Energy Conversion"}
@STRING{IEEE_J_PWRAS      = "{IEEE} Trans. Power App. Syst."}
@STRING{IEEE_J_PWRD       = "{IEEE} Trans. Power Delivery"}
@STRING{IEEE_J_PWRE       = "{IEEE} Trans. Power Electron."}
@STRING{IEEE_J_PWRS       = "{IEEE} Trans. Power Syst."}



@STRING{IEEE_J_APPIND     = "{IEEE} Trans. Applicat. Ind."}
@STRING{IEEE_J_BC         = "{IEEE} Trans. Broadcast."}
STRING{IEEE_J_BCTV        = "{IEEE} Trans. Broadcast Television Receivers"}
@STRING{IEEE_J_CE         = "{IEEE} Trans. Consumer Electron."}
@STRING{IEEE_J_IE         = "{IEEE} Trans. Ind. Electron."}
@STRING{IEEE_J_IECI       = "{IEEE} Trans. Ind. Electron. Contr. Instrum."}
@STRING{IEEE_J_IA         = "{IEEE} Trans. Ind. Applicat."}
@STRING{IEEE_J_IGA        = "{IEEE} Trans. Ind. Gen. Applicat."}



@STRING{IEEE_J_IM         = "{IEEE} Trans. Instrum. Meas."}



STRING{IEEE_J_JEM         = "{IEEE/TMS} J. Electron. Mater."}
@STRING{IEEE_J_DEI        = "{IEEE} Trans. Dielect. Elect. Insulation"}
@STRING{IEEE_J_EI         = "{IEEE} Trans. Elect. Insulation"}



@STRING{IEEE_J_MECH       = "{IEEE/ASME} Trans. Mechatron."}
@STRING{IEEE_J_MEMS       = "J. Microelectromech. Syst."}



@STRING{IEEE_J_BME        = "{IEEE} Trans. Biomed. Eng."}
@STRING{IEEE_J_B-ME       = "{IEEE} Trans. Bio-Med. Eng."}
@STRING{IEEE_J_BMELC      = "{IEEE} Trans. Bio-Med. Electron."}
@STRING{IEEE_J_ITBM       = "{IEEE} Trans. Inform. Technol. Biomed."}
@STRING{IEEE_J_ME         = "{IEEE} Trans. Med. Electron."}
@STRING{IEEE_J_MI         = "{IEEE} Trans. Med. Imag."}
STRING{IEEE_J_MCTE        = "{IEEE} Trans. Molecular Cellular Tissue Eng."}
@STRING{IEEE_J_NB         = "{IEEE} Trans. Nanobiosci."}
@STRING{IEEE_J_NSRE       = "{IEEE} Trans. Neural Syst. Rehab. Eng."}
@STRING{IEEE_J_RE         = "{IEEE} Trans. Rehab. Eng."}



@STRING{IEEE_J_PTL        = "{IEEE} Photon. Technol. Lett."}
@STRING{IEEE_J_JLT        = "J. Lightwave Technol."}



@STRING{IEEE_J_EDL        = "{IEEE} Electron Device Lett."}
@STRING{IEEE_J_JQE        = "{IEEE} J. Quantum Electron."}
@STRING{IEEE_J_JSTQE      = "{IEEE} J. Select. Topics Quantum Electron."}
@STRING{IEEE_J_ED         = "{IEEE} Trans. Electron Devices"}
@STRING{IEEE_J_NANO       = "{IEEE} Trans. Nanotechnol."}
@STRING{IEEE_J_NS         = "{IEEE} Trans. Nucl. Sci."}
@STRING{IEEE_J_PS         = "{IEEE} Trans. Plasma Sci."}



@STRING{IEEE_J_DMR        = "{IEEE} Trans. Device Mat. Rel."}
@STRING{IEEE_J_R          = "{IEEE} Trans. Rel."}



STRING{IEEE_J_ESSL        = "{IEEE/ECS} Electrochemical Solid-State Lett."}
@STRING{IEEE_J_JSSC       = "{IEEE} J. Solid-State Circuits"}
@STRING{IEEE_J_ASC        = "{IEEE} Trans. Appl. Superconduct."}
@STRING{IEEE_J_SM         = "{IEEE} Trans. Semiconduct. Manufact."}



@STRING{IEEE_J_SENSOR     = "{IEEE} Sensors J."}



@STRING{IEEE_J_VLSI       = "{IEEE} Trans. {VLSI} Syst."}







@STRING{IEEE_M_AES        = "{IEEE} Aerosp. Electron. Syst. Mag"}
@STRING{IEEE_M_HIST       = "{IEEE} Annals Hist. Comput."}
@STRING{IEEE_M_AP         = "{IEEE} Antennas Propagat. Mag."}
@STRING{IEEE_M_ASSP       = "{IEEE} {ASSP} Mag."}
@STRING{IEEE_M_CD         = "{IEEE} Circuits Devices Mag."}
@STRING{IEEE_M_CAS        = "{IEEE} Circuits Syst. Mag."}
@STRING{IEEE_M_COM        = "{IEEE} Commun. Mag."}
@STRING{IEEE_M_COMSOC     = "{IEEE} Commun. Soc. Mag."}
@STRING{IEEE_M_CSE        = "{IEEE} Comput. Sci. Eng."}
@STRING{IEEE_M_CSEM       = "{IEEE} Comput. Sci. Eng. Mag."}
@STRING{IEEE_M_C          = "{IEEE} Computer"}
@STRING{IEEE_M_CAP        = "{IEEE} Comput. Appl. Power"}
@STRING{IEEE_M_CGA        = "{IEEE} Comput. Graph. Appl."}
@STRING{IEEE_M_CONC       = "{IEEE} Concurrency"}
@STRING{IEEE_M_CS         = "{IEEE} Control Syst. Mag."}
@STRING{IEEE_M_DTC        = "{IEEE} Des. Test. Comput."}
@STRING{IEEE_M_EI         = "{IEEE} Electr. Insul. Mag."}
@STRING{IEEE_M_EMB        = "{IEEE} Eng. Med. Biol. Mag."}
@STRING{IEEE_M_EMR        = "{IEEE} Eng. Manag. Rev."}
@STRING{IEEE_M_EXP        = "{IEEE} Expert"}
@STRING{IEEE_M_IA         = "{IEEE} Ind. Appl. Mag."}
@STRING{IEEE_M_IM         = "{IEEE} Instrum. Meas. Mag."}
@STRING{IEEE_M_IS         = "{IEEE} Intell. Syst."}
@STRING{IEEE_M_IC         = "{IEEE} Internet Comput."}
@STRING{IEEE_M_ITP        = "{IEEE} {IT} Prof."}
@STRING{IEEE_M_MICRO      = "{IEEE} Micro"}
@STRING{IEEE_M_MW         = "{IEEE} Microwave"}
@STRING{IEEE_M_MM         = "{IEEE} Multimedia"}
@STRING{IEEE_M_NET        = "{IEEE} Network"}
@STRING{IEEE_M_PCOM       = "{IEEE} Personal Commun. Mag."}
@STRING{IEEE_M_POT        = "{IEEE} Potentials"}
@STRING{IEEE_M_PE         = "{IEEE} Power Energy Mag."}
@STRING{IEEE_M_PER        = "{IEEE} Power Eng. Rev."}
@STRING{IEEE_M_RA         = "{IEEE} Robot. Automat. Mag."}
@STRING{IEEE_M_SP         = "{IEEE} Signal Processing Mag."}
@STRING{IEEE_M_S          = "{IEEE} Softw."}
@STRING{IEEE_M_SPECT      = "{IEEE} Spectr."}
@STRING{IEEE_M_TS         = "{IEEE} Technol. Soc. Mag."}
@STRING{IEEE_M_WC         = "{IEEE} Wireless Commun. Mag."}
@STRING{IEEE_M_TODAY      = "Today's Eng."}






%
%
%


@STRING{acmcs     = "ACM Computing Surveys"}
@STRING{acta      = "Acta Informatica"}
@STRING{cacm      = "Communications of the ACM"}
@STRING{ibmjrd    = "IBM Journal of Research and Development"}
@STRING{ibmsj     = "IBM Systems Journal"}
@STRING{ieeese    = "IEEE Transactions on Software Engineering"}
@STRING{ieeetc    = "IEEE Transactions on Computers"}
@STRING{ieeetcad  = "IEEE Transactions on Computer-Aided Design of Integrated Circuits"}
@STRING{ipl       = "Information Processing Letters"}
@STRING{jacm      = "Journal of the ACM"}
@STRING{jcss      = "Journal of Computer and System Sciences"}
@STRING{scp       = "Science of Computer Programming"}
@STRING{sicomp    = "SIAM Journal on Computing"}
@STRING{tocs      = "ACM Transactions on Computer Systems"}
@STRING{tods      = "ACM Transactions on Database Systems"}
@STRING{tog       = "ACM Transactions on Graphics"}
@STRING{toms      = "ACM Transactions on Mathematical Software"}
@STRING{toois     = "ACM Transactions on Office Information Systems"}
@STRING{toplas    = "ACM Transactions on Programming Languages and Systems"}
@STRING{tcs       = "Theoretical Computer Science"}








 
 
 
 
@electronic{IEEEexample:shellCTANpage,
  author        = "Michael Shell",
  title         = "{IEEE}tran Homepage on {CTAN}",
  url           = "http://www.ctan.org/tex-archive/macros/latex/contrib/supported/IEEEtran/",
  year          = "2002"
};

@electronic{IEEEexample:IEEEwebsite,
  title         = "The {IEEE} Website",
  url           = "http://www.ieee.org/",
  year          = "2002",
  key           = "IEEE"
};

@electronic{IEEEexample:bibtexuser,
  author        = "Oren Patashnik",
  title         = "{\BibTeX}ing",
  howpublished  = "{btxdoc.pdf}",
  url           = "http://www.ctan.org/tex-archive/biblio/bibtex/contrib/doc/",
  month         = feb,
  year          = "1988"
};

@electronic{IEEEexample:bibtexdesign,
  author        = "Oren Patashnik",
  title         = "Designing {\BibTeX\ } Styles",
  howpublished  = "{btxhak.pdf}",
  url           = "http://www.ctan.org/tex-archive/biblio/bibtex/contrib/doc/",
  month         = feb,
  year          = "1988"
};

@electronic{IEEEexample:bibtexFAQ,
  author        = "David Hoadley and Michael Shell",
  title         = "{\BibTeX}\ Tips and {FAQ}",
  howpublished  = "{btxFAQ.txt}",
  url           = "http://www.ctan.org/tex-archive/biblio/bibtex/contrib/doc/",
  month         = oct,
  year          = "2002"
};

@electronic{IEEEexample:beebe_archive,
  author        = "Nelson H. F. Beebe",
  title         = "{\TeX\ }User Group Bibliography Archive",
  url           = "http://www.math.utah.edu:8080/pub/tex/bib/index-table.html",
  month         = may,
  year          = "2002"
};

@electronic{IEEEexample:urlsty,
  author        = "Donald Arseneau",
  title         = "The url.sty package",
  url           = "http://www.ctan.org/tex-archive/macros/latex/contrib/other/misc/",
  month         = mar,
  year          = "1999",
};


@electronic{IEEEexample:hyperrefsty,
  author        = "Sebastian Rahtz and Heiko Oberdiek",
  title         = "The hyperref.sty package",
  url           = "http://www.ctan.org/tex-archive/macros/latex/contrib/supported/hyperref/",
  month         = jul,
  year          = "2002",
};

@electronic{IEEEexample:babel,
  author        = "Johannes Braams",
  title         = "The {Babel} package",
  url           = "http://www.ctan.org/tex-archive/macros/latex/required/babel/",
  month         = feb,
  year          = "2001",
};






@article{IEEEexample:article_typical,
  author        = "S. Zhang and C. Zhu and J. K. O. Sin and P. K. T. Mok",
  title         = "A Novel Ultrathin Elevated Channel Low-temperature 
                   Poly-{Si} {TFT}",
  journal       = IEEE_J_EDL,
  volume        = "20",
  month         = nov,
  year          = "1999",
  pages         = "569-571"
};

@article{IEEEexample:articleetal,
  author        = "F. Delorme and others",
  title         = "Butt-jointed {DBR} Laser With 15 {nm} Tunability Grown
                   in Three {MOVPE} Steps",
  journal       = "Electron. Lett.",
  volume        = "31",
  number        = "15",
  year          = "1995",
  pages         = "1244-1245"
};


@inproceedings{IEEEexample:conf_typical,
  author        = "R. K. Gupta and S. D. Senturia",
  title         = "Pull-in Time Dynamics as a Measure of Absolute Pressure",
  booktitle     = "Proc. {IEEE} International Workshop on
                   Microelectromechanical Systems ({MEMS}'97)",
  address       = "Nagoya, Japan",
  month         = jan,
  year          = "1997",
  pages         = "290-294"
};


@book{IEEEexample:book_typical,
  author        = "B. D. Cullity",
  title         = "Introduction to Magnetic Materials",
  publisher     = "Addison-Wesley",
  address       = "Reading, MA",
  year          = "1972"
};





@article{IEEEexample:articlelargepages,
  author        = "A. Castaldini and A. Cavallini and B. Fraboni
                   and P. Fernandez and J. Piqueras",
  title         = "Midgap Traps Related to Compensation Processes in
                   {CdTe} Alloys",
  journal       = "Phys. Rev. B.",
  volume        = "56",
  number        = "23",
  year          = "1997",
  pages         = "14897-14900"
};


@article{IEEEexample:articledualmonths,
  author        = "Y. Okada and K. Dejima and T. Ohishi",
  title         = "Analysis and Comparison of {PM} Synchronous Motor and
                   Induction Motor Type Magnetic Bearings",
  journal       = IEEE_J_IA,
  volume        = "31",
  month         = sep # "/" # oct,
  year          = "1995",
  pages         = "1047-1053"
};


@misc{IEEEexample:TBPmisc,
  author        = "M. Coates and A. Hero and R. Nowak and B. Yu",
  title         = "Internet Tomography",
  howpublished  = IEEE_M_SP,
  month         = may,
  year          = "2002",
  note          = "to be published"
};


@article{IEEEexample:TBParticle,
  author        = "N. Kahale and R. Urbanke",
  title         = "On the Minimum Distance of Parallel and Serially
                   Concatenated Codes",
  journal       = IEEE_J_IT,
  year          = "submitted for publication"
};





@book{IEEEexample:bookwitheditor,
  editor        = "J. C. Candy and G. C. Temes",
  title         = "Oversampling Delta-Sigma Data Converters Theory,
                   Design and Simulation",
  publisher     = "{IEEE} Press.",
  address       = "New York",
  year          = "1992"
};


@book{IEEEexample:book,
  author        = "S. M. Metev and V. P. Veiko",
  editor        = "Osgood, Jr., R. M.",
  title         = "Laser Assisted Microtechnology",
  edition       = "Second",
  publisher     = "Springer-Verlag",
  address       = "Berlin, Germany",
  year          = "1998"
};


@book{IEEEexample:bookwithseriesvolume,
  editor        = "J. Breckling",
  title         = "The Analysis of Directional Time Series: Applications to
                   Wind Speed and Direction",
  series        = "Lecture Notes in Statistics",
  publisher     = "Springer",
  address       = "Berlin, Germany",
  year          = "1989",
  volume        = "61"
};


@inbook{IEEEexample:inbook,
  author        = "H. E. Rose",
  title         = "A Course in Number Theory",
  publisher     = "Oxford Univ. Press",
  address       = "New York, NY",
  year          = "1988",
  chapter       = "3"
};


@inbook{IEEEexample:inbookpagesnote,
  author        = "B. K. Bul",
  title         = "Theory Principles and Design of Magnetic Circuits",
  publisher     = "Energia Press",
  address       = "Moscow",
  year          = "1964",
  pages         = "464",
  note          = "(in Russian)"
};





@incollection{IEEEexample:incollection,
  author        = "W. V. Sorin",
  editor        = "D. Derickson",
  title         = "Optical Reflectometry for Component Characterization",
  booktitle     = "Fiber Optic Test and Measurement",
  publisher     = "Prentice-Hall",
  address       = "Englewood Cliffs, NJ",
  year          = "1998"
};


@incollection{IEEEexample:incollectionwithseries,
  author        = "J. B. Anderson and K. Tepe",
  title         = "Properties of the Tailbiting {BCJR} Decoder",
  booktitle     = "Codes, Systems and Graphical Models",
  series        = "{IMA} Volumes in Mathematics and Its Applications",
  publisher     = "Springer-Verlag",
  address       = "New York",
  year          = "2000"
  
};


@incollection{IEEEexample:incollection_chpp,
  author        = "P. Hedelin and P. Knagenhjelm and M. Skoglund",
  editor        = "W. B. Kleijn and K. K. Paliwal",
  title         = "Theory for Transmission of Vector Quantization Data",
  booktitle     = "Speech Coding and Synthesis",
  publisher     = "Elsevier Science",
  address       = "Amsterdam, The Netherlands",
  year          = "1995",
  chapter       = "10",
  pages         = "347-396"
};


@incollection{IEEEexample:incollectionmanyauthors,
  author        = "R. M. A. Dawson and Z. Shen and D. A. Furst and
                   S. Connor and J. Hsu and M. G. Kane and R. G. Stewart and
                   A. Ipri and C. N. King and P. J. Green and R. T. Flegal
                   and S. Pearson and W. A. Barrow and E. Dickey and K. Ping
                   and C. W. Tang and S. Van. Slyke and
                   F. Chen and J. Shi and J. C. Sturm and M. H. Lu",
  title         = "Design of an Improved Pixel for a Polysilicon 
                   Active-Matrix Organic {LED} Display",
  booktitle     = "{SID} Tech. Dig.",
  volume        = "29",
  year          = "1998",
  pages         = "11-14"
};





@manual{IEEEexample:motmanual,
  title         = "{FLEXChip} Signal Processor ({MC68175/D})",
  organization  = "Motorola",
  year          = "1996"
};

@manual{IEEEexample:motmanualhowpub,
  title         = "{FLEXChip} Signal Processor",
  howpublished  = "{MC68175/D}",
  organization  = "Motorola",
  year          = "1996"
};




@inproceedings{IEEEexample:confwithadddays,
  author        = "M. S. Yee and L. Hanzo",
  title         = "Radial Basis Function Decision Feedback Equaliser
                   Assisted Burst-by-burst Adaptive Modulation",
  booktitle     = "Proc. {IEEE} Globecom '99",
  address       = "Rio de Janeiro, Brazil",
  month         = dec # " 5--9,",
  year          = "1999",
  pages         = "2183-2187"
};


@inproceedings{IEEEexample:confwithvolume,
  author        = "M. Yajnik and S. B. Moon and J. Kurose and D. Towsley",
  title         = "Measurement and Modeling of the Temporal Dependence in
                   Packet Loss",
  booktitle     = "Proc. {IEEE} {INFOCOM}'99",
  volume        = "1",
  address       = "New York, NY",
  month         = mar,
  year          = "1999",
  pages         = "345-352"
};


@inproceedings{IEEEexample:confwithpaper,
  author        = "M. Wegmuller and J. P. von der Weid and P. Oberson
                   and N. Gisin",
  title         = "High Resolution Fiber Distributed Measurements With
                   Coherent {OFDR}",
  booktitle     = "Proc. {ECOC}'00",
  year          = "2000",
  paper         = "11.3.4",
  pages         = "109"
};


@inproceedings{IEEEexample:confwithpapertype,
  author        = "B. Mikkelsen and G. Raybon and R.-J. Essiambre and
                   K. Dreyer and Y. Su. and L. E. Nelson and J. E. Johnson
                   and G. Shtengel and A. Bond and D. G. Moodie and
                   A. D. Ellis",
  title         = "160 {Gbit/s} Single-channel Transmission Over 300 km 
                   Nonzero-dispersion Fiber With Semiconductor Based
                   Transmitter and Demultiplexer",
  booktitle     = "Proc. {ECOC}'99",
  year          = "1999",
  paper         = "2-3",
  type          = "postdeadline paper",
  pages         = "28-29"
};


@inproceedings{IEEEexample:presentedatconf,
  author        = "S. G. Finn and M. M{\'e}dard and R. A. Barry",
  title         = "A Novel Approach to Automatic Protection Switching
                   Using Trees",
  intype        = "presented at the",
  booktitle     = "Proc. Int. Conf. Commun.",
  year          = "1997"
};





@mastersthesis{IEEEexample:masters,
  author        = "Nin C. Loh",
  title         = "High-Resolution Micromachined Interferometric
                   Accelerometer",
  school        = "Massachusetts Institute of Technology",
  address       = "Cambridge",
  year          = "1992"
};


@mastersthesis{IEEEexample:masterstype,
  author        = "A. Karnik",
  title         = "Performance of {TCP} Congestion Control with Rate
                   Feedback: {TCP/ABR} and Rate Adaptive {TCP/IP}",
  school        = "Indian Institute of Science",
  type          = "M. Eng. thesis",
  address       = "Bangalore, India",
  month         = jan,
  year          = "1999"
};





@phdthesis{IEEEexample:phdurl,
  author        = "Q. Li",
  title         = "Delay Characterization and Performance Control of
                   Wide-area Networks",
  school        = "Univ. of Delaware",
  address       = "Newark",
  month         = may,
  year          = "2000",
  url           = "http://www.ece.udel.edu/~qli"
};





@techreport{IEEEexample:techrep,
  author        = "R. Jain and K. K. Ramakrishnan and D. M. Chiu",
  title         = "Congestion Avoidance in Computer Networks with a 
                   Connectionless Network Layer",
  institution   = "Digital Equipment Corporation",
  address       = "MA",
  number        = "DEC-TR-506",
  month         = aug,
  year          = "1987"
};


@techreport{IEEEexample:techreptype,
  author        = "J. Padhye and V. Firoiu and D. Towsley",
  title         = "A Stochastic Model of {TCP} {R}eno Congestion Avoidance
                   and Control",
  institution   = "Univ. of Massachusetts",
  address       = "Amherst, MA",
  type          = "CMPSCI Tech. Rep.",
  number        = "99-02",
  year          = "1999"
};


@techreport{IEEEexample:techreptypeii,
  author        = "D. Middleton and A. D. Spaulding",
  title         = "A Tutorial Review of Elements of Weak Signal Detection
                   in Non-{G}aussian {EMI} Environments",
  institution   = "National Telecommunications and Information
                   Administration ({NTIA}), U.S. Dept. of Commerce",
  type          = "NTIA Report",
  number        = "86-194",
  month         = may,
  year          = "1986"
};





@unpublished{IEEEexample:unpublished,
  author        = "T. J. Ott and N. Aggarwal",
  title         = "{TCP} over {ATM}: {ABR} or {UBR}",
  note          = "Unpublished"
};





@electronic{IEEEexample:electronhowinfo,
  author        = "V. Jacobson",
  title         = "Modified {TCP} Congestion Avoidance Algorithm",
  howpublished  = "end2end-interest mailing list",
  url           = "ftp://ftp.isi.edu/end2end/end2end-interest-1990.mail",
  month         = apr,
  year          = "1990"
};


@electronic{IEEEexample:electronhowinfo2,
  author        = "V. Valloppillil and K. W. Ross",
  title         = "Cache Array Routing Protocol v1.1",
  howpublished  = "Internet draft",
  url           = "http://ds1.internic.net/internet-drafts/draft-vinod-carp-v1-03.txt",
  year          = "1998"
};


@electronic{IEEEexample:electronorgadd,
  author        = "D. H. Lorenz and A. Orda",
  title         = "Optimal Partition of {QoS} Requirements on Unicast
                   Paths and Multicast Trees",
  organization  = "Dept. Elect. Eng., Technion",
  address       = "Haifa, Israel",
  url           = "ftp://ftp.technion.ac.il/pub/supported/ee/Network/lor.mopq98.ps",
  month         = jul,
  year          = "1998"
};





@patent{IEEEexample:uspat,
  author        = "Ronald E. Sorace and Victor S. Reinhardt and
                   Steven A. Vaughn",
  assignee      = "Hughes Aircraft Company",
  address       = "Los Angeles, CA",
  title         = "High-Speed Digital-to-{RF} Converter",
  nationality   = "United States",
  number        = "5668842",
  dayfiled      = "28",
  monthfiled    = feb,
  yearfiled     = "1995",
  day           = "16",
  month         = sep,
  year          = "1997"
};


@patent{IEEEexample:jppat,
  author        = "U. Hideki",
  title         = "Quadrature Modulation Circuit",
  nationality   = "Japanese",
  number        = "152932/92",
  day           = "20",
  month         = may,
  year          = "1992"
};


@patent{IEEEexample:frenchpatreq,
  author        = "F. Kowalik and M. Isard",
  title         = "Estimateur d'un D{\'e}faut de Fonctionnement 
                   d'un Modulateur en Quadrature et {\'E}tage de Modulation
                   l'Utilisant",
  language      = "french",
  nationality   = "French",
  type          = "Patent Request",
  number        = "9500261",
  dayfiled      = "11",
  monthfiled    = jan,
  yearfiled     = "1995"
};





@periodical{IEEEexample:periodical,
  title         = IEEE_M_PCOM # ", Special Issue on Wireless {ATM}",
  volume        = "3",
  month         = aug,
  year          = "1996",
  key           = "IEEE"
};





@standard{IEEEexample:standard,
  title         = "Wireless {LAN} Medium Access Control {(MAC)} and 
                   Physical Layer {(PHY)} Specification",
  organization  = "IEEE",
  address       = "Piscataway, NJ",
  number        = "802.11",
  year          = "1997"
};


@standard{IEEEexample:standardproposed,
  title         = "Fiber Channel Physical Interface ({FC-PI})",
  institution   = "NCITS",
  address       = "Washington, DC",
  type          = "Working Draft Proposed Standard",
  revision      = "5.2",
  year          = "1999"
};


@misc{IEEEexample:draftasmisc,
  author        = "I. Widjaja and A. Elwalid",
  title         = "{MATE}: {MPLS} Adaptive Traffic Engineering",
  howpublished  = "IETF Draft",
  year          = "1999"
};





@misc{IEEEexample:miscforum,
  author        = "L. Roberts",
  title         = "Enhanced Proportional Rate Control Algorithm {PRCA}",
  howpublished  = "{ATM} Forum Contribution 94-0735R1",
  month         = aug,
  year          = "1994"
};


@misc{IEEEexample:whitepaper,
  title         = "Advanced {QoS} Services for the Intelligent Internet",
  howpublished  = "White Paper",
  organization  = "Cisco",
  month         = may,
  year          = "1997"
};


@misc{IEEEexample:datasheet,
  title         = "{PDCA12-70} data sheet",
  organization  = "Opto Speed SA",
  address       = "Mezzovico, Switzerland"
};






@misc{IEEEexample:private,
  author        = "S. Konyagin",
  howpublished  = "private communication",
  year          = "1998"
};


@misc{IEEEexample:miscrfc,
  author        = "K. K. Ramakrishnan and S. Floyd",
  title         = "A Proposal to Add Explicit Congestion
                   Notification ({ECN}) to {IP}",
  howpublished  = "RFC 2481",
  month         = jan,
  year          = "1999"
};


@misc{IEEEexample:miscgermanreg,
  title         = "{M}essung von {S}t{\"o}rfeldern an {A}nlagen 
                   und {L}eitungen der {T}elekommunikation im
                   {F}requenzbereich 9 {kHz} bis 3 {GHz}",
  language      = "german",
  howpublished  = "{M}e{\ss}vorschrift {R}eg {TP} {MV} 05",
  organization  = "Regulierungsbeh{\"o}rde f{\"u}r {T}elekommunikation und
                   {P}ost ({R}eg {TP})"
};



@article{IEEEexample:bluebookarticle,
  author        = "{Consulative Committee for Space Data Systems (CCSDS)}",
  title         = "Telemetry Channel Coding",
  journal       = "Blue Book",
  number        = "4",
  year          = "1999",
  url           = "http://www.ccsds.org/documents/pdf/CCSDS-101.0-B-4.pdf"
};


@book{IEEEexample:bluebookbook,
  author        = "{Consulative Committee for Space Data Systems (CCSDS)}",
  title         = "Telemetry Channel Coding",
  series        = "Blue Book",
  number        = "4",
  publisher     = "CCSDS",
  address       = "Newport Beach, CA",
  year          = "1999",
  url           = "http://www.ccsds.org/documents/pdf/CCSDS-101.0-B-4.pdf"
};

@manual{IEEEexample:bluebookmanual,
  title         = "Telemetry Channel Coding",
  howpublished  = "ser. Blue Book, No. 4",
  organization  = "Consulative Committee for Space Data Systems (CCSDS)",
  address       = "Newport Beach, CA",
  year          = "1999",
  url           = "http://www.ccsds.org/documents/pdf/CCSDS-101.0-B-4.pdf"
};



@standard{IEEEexample:bluebookstandard,
  title         = "Telemetry Channel Coding",
  howpublished  = "ser. Blue Book, No. 4",
  organization  = "Consulative Committee for Space Data Systems (CCSDS)",
  address       = "Newport Beach, CA",
  type          = "Recommendation for Space Data System Standard",
  number        = "101.0-B-4",
  month         = May,
  year          = "1999",
  url           = "http://www.ccsds.org/documents/pdf/CCSDS-101.0-B-4.pdf"
};





 
%
%
%
%
%
%
%
%
%
%
%
@IEEEtranBSTCTL{IEEEexample:BSTcontrol,
  CTLuse_article_number     = "yes",
  CTLuse_paper              = "yes",
  CTLuse_forced_etal        = "no",
  CTLmax_names_forced_etal  = "10",
  CTLnames_show_etal        = "1",
  CTLuse_alt_spacing        = "yes",
  CTLalt_stretch_factor     = "4",
  CTLdash_repeated_names    = "yes",
  CTLname_format_string     = "{f.~}{vv~}{ll}{, jj}",
  CTLname_latex_cmd         = ""
};





%
%
%
%
%
%
%
%
%
%
%
%
%
%
%
%







@STRING{IEEE_J_AES        = "{IEEE} Transactions on Aerospace and Electronic Systems"}
@STRING{IEEE_J_ANE        = "{IEEE} Transactions on Aerospace and Navigational Electronics"}
@STRING{IEEE_J_ANNE       = "{IEEE} Transactions on Aeronautical and Navigational Electronics"}
@STRING{IEEE_J_AS         = "{IEEE} Transactions on Aerospace"}
@STRING{IEEE_J_AIRE       = "{IEEE} Transactions on Airborne Electronics"}
@STRING{IEEE_J_MIL        = "{IEEE} Transactions on Military Electronics"}



@STRING{IEEE_J_ITS        = "{IEEE} Transactions on Intelligent Transportation Systems"}
@STRING{IEEE_J_VT         = "{IEEE} Transactions on Vehicular Technology"}
@STRING{IEEE_J_VC         = "{IEEE} Transactions on Vehicular Communications"}



@STRING{IEEE_J_SPL        = "{IEEE} Signal Processing Letters"}
@STRING{IEEE_J_ASSP       = "{IEEE} Transactions on Acoustics, Speech, and Signal Processing"}
@STRING{IEEE_J_AU         = "{IEEE} Transactions on Audio"}
@STRING{IEEE_J_AUEA       = "{IEEE} Transactions on Audio and Electroacoustics"}
@STRING{IEEE_J_AC         = "{IEEE} Transactions on Automatic Control"}
@STRING{IEEE_J_CAS        = "{IEEE} Transactions on Circuits and Systems"}
@STRING{IEEE_J_CASVT      = "{IEEE} Transactions on Circuits and Systems for Video Technology"}
@STRING{IEEE_J_CASI       = "{IEEE} Transactions on Circuits and Systems---Part {I}: Fundamental Theory and Applications"}
@STRING{IEEE_J_CASII      = "{IEEE} Transactions on Circuits and Systems---Part {II}: Analog and Digital Signal Processing"}
@STRING{IEEE_J_CT         = "{IEEE} Transactions on Circuit Theory"}
@STRING{IEEE_J_CST        = "{IEEE} Transactions on Control Systems Technology"}
@STRING{IEEE_J_SP         = "{IEEE} Transactions on Signal Processing"}
@STRING{IEEE_J_SU         = "{IEEE} Transactions on Sonics and Ultrasonics"}
@STRING{IEEE_J_SAP        = "{IEEE} Transactions on Speech and Audio Processing"}
@STRING{IEEE_J_UE         = "{IEEE} Transactions on Ultrasonics Engineering"}
@STRING{IEEE_J_UFFC       = "{IEEE} Transactions on Ultrasonics, Ferroelectrics, and Frequency Control"}



@STRING{IEEE_J_COML       = "{IEEE} Communications Letters"}
@STRING{IEEE_J_JSAC       = "{IEEE} Journal on Selected Areas in Communications"}
@STRING{IEEE_J_COM        = "{IEEE} Transactions on Communications"}
@STRING{IEEE_J_COMT       = "{IEEE} Transactions on Communication Technology"}
@STRING{IEEE_J_WCOM       = "{IEEE} Transactions on Wireless Communications"}



@STRING{IEEE_J_ADVP       = "{IEEE} Transactions on Advanced Packaging"}
@STRING{IEEE_J_CHMT       = "{IEEE} Transactions on Components, Hybrids and Manufacturing Technology"}
@STRING{IEEE_J_CPMTA      = "{IEEE} Transactions on Components, Packaging and Manufacturing Technology---Part {A}"}
@STRING{IEEE_J_CPMTB      = "{IEEE} Transactions on Components, Packaging and Manufacturing Technology---Part {B}: Advanced Packaging"}
@STRING{IEEE_J_CPMTC      = "{IEEE} Transactions on Components, Packaging and Manufacturing Technology---Part {C}: Manufacturing"}
@STRING{IEEE_J_CAPT       = "{IEEE} Transactions on Components and Packaging Technology"}
@STRING{IEEE_J_CAPTS      = "{IEEE} Transactions on Components and Packaging Technologies"}
@STRING{IEEE_J_CPART      = "{IEEE} Transactions on Component Parts"}
@STRING{IEEE_J_EPM        = "{IEEE} Transactions on Electronics Packaging Manufacturing"}
@STRING{IEEE_J_MFT        = "{IEEE} Transactions on Manufacturing Technology"}
@STRING{IEEE_J_PHP        = "{IEEE} Transactions on Parts, Hybrids and Packaging"}
@STRING{IEEE_J_PMP        = "{IEEE} Transactions on Parts, Materials and Packaging"}



@STRING{IEEE_J_TCAD       = "{IEEE} Journal on Technology in Computer Aided Design"}
@STRING{IEEE_J_CAD        = "{IEEE} Transactions on Computer-Aided Design of Integrated Circuits and Systems"}



@STRING{IEEE_J_IT         = "{IEEE} Transactions on Information Theory"}
@STRING{IEEE_J_KDE        = "{IEEE} Transactions on Knowledge and Data Engineering"}



@STRING{IEEE_J_C          = "{IEEE} Transactions on Computers"}
@STRING{IEEE_J_ECOMP      = "{IEEE} Transactions on Electronic Computers"}
@STRING{IEEE_J_EVC        = "{IEEE} Transactions on Evolutionary Computation"}
@STRING{IEEE_J_FUZZ       = "{IEEE} Transactions on Fuzzy Systems"}
@STRING{IEEE_J_MC         = "{IEEE} Transactions on Mobile Computing"}
@STRING{IEEE_J_NET        = "{IEEE/ACM} Transactions on Networking"}
@STRING{IEEE_J_NN         = "{IEEE} Transactions on Neural Networks"}
@STRING{IEEE_J_PDS        = "{IEEE} Transactions on Parallel and Distributed Systems"}
@STRING{IEEE_J_SE         = "{IEEE} Transactions on Software Engineering"}



@STRING{IEEE_J_IP         = "{IEEE} Transactions on Image Processing"}
@STRING{IEEE_J_MM         = "{IEEE} Transactions on Multimedia"}
@STRING{IEEE_J_VCG        = "{IEEE} Transactions on Visualization and Computer Graphics"}



@STRING{IEEE_J_JRA        = "{IEEE} Journal of Robotics and Automation"}
@STRING{IEEE_J_HFE        = "{IEEE} Transactions on Human Factors in Electronics"}
@STRING{IEEE_J_MMS        = "{IEEE} Transactions on Man-Machine Systems"}
@STRING{IEEE_J_PAMI       = "{IEEE} Transactions on Pattern Analysis and Machine Intelligence"}
@STRING{IEEE_J_RA         = "{IEEE} Transactions on Robotics and Automation"}
@STRING{IEEE_J_SMC        = "{IEEE} Transactions on Systems, Man, and Cybernetics"}
@STRING{IEEE_J_SMCA       = "{IEEE} Transactions on Systems, Man, and Cybernetics---Part {A}: Systems and Humans"}
@STRING{IEEE_J_SMCB       = "{IEEE} Transactions on Systems, Man, and Cybernetics---Part {B}: Cybernetics"}
@STRING{IEEE_J_SMCC       = "{IEEE} Transactions on Systems, Man, and Cybernetics---Part {C}: Applications and Reviews"}
@STRING{IEEE_J_SSC        = "{IEEE} Transactions on Systems Science and Cybernetics"}



@STRING{IEEE_J_GE         = "{IEEE} Transactions on Geoscience Electronics"}
@STRING{IEEE_J_GRS        = "{IEEE} Transactions on Geoscience and Remote Sensing"}
@STRING{IEEE_J_OE         = "{IEEE} Journal of Oceanic Engineering"}



@STRING{IEEE_J_CJECE      = "Canadian Journal of Electrical and Computer Engineering"}
@STRING{IEEE_J_PROC       = "Proceedings of the {IEEE}"}
@STRING{IEEE_J_EDU        = "{IEEE} Transactions on Education"}
@STRING{IEEE_J_EM         = "{IEEE} Transactions on Engineering Management"}
@STRING{IEEE_J_EWS        = "{IEEE} Transactions on Engineering Writing and Speech"}
@STRING{IEEE_J_PC         = "{IEEE} Transactions on Professional Communication"}



@STRING{IEEE_J_AWPL       = "{IEEE} Antennas and Wireless Propagation Letters"}
@STRING{IEEE_J_MGWL       = "{IEEE} Microwave and Guided Wave Letters"}
@STRING{IEEE_J_MWCL       = "{IEEE} Microwave and Wireless Components Letters"}
@STRING{IEEE_J_AP         = "{IEEE} Transactions on Antennas and Propagation"}
@STRING{IEEE_J_EMC        = "{IEEE} Transactions on Electromagnetic Compatibility"}
@STRING{IEEE_J_MAG        = "{IEEE} Transactions on Magnetics"}
@STRING{IEEE_J_MTT        = "{IEEE} Transactions on Microwave Theory and Techniques"}
@STRING{IEEE_J_RFI        = "{IEEE} Transactions on Radio Frequency Interference"}
@STRING{IEEE_J_TJMJ       = "{IEEE} Translation Journal on Magnetics in Japan"}



@STRING{IEEE_J_EC         = "{IEEE} Transactions on Energy Conversion"}
@STRING{IEEE_J_PWRAS      = "{IEEE} Transactions on Power Apparatus and Systems"}
@STRING{IEEE_J_PWRD       = "{IEEE} Transactions on Power Delivery"}
@STRING{IEEE_J_PWRE       = "{IEEE} Transactions on Power Electronics"}
@STRING{IEEE_J_PWRS       = "{IEEE} Transactions on Power Systems"}



@STRING{IEEE_J_APPIND     = "{IEEE} Transactions on Applications and Industry"}
@STRING{IEEE_J_BC         = "{IEEE} Transactions on Broadcasting"}
@STRING{IEEE_J_BCTV       = "{IEEE} Transactions on Broadcast and Television Receivers"}
@STRING{IEEE_J_CE         = "{IEEE} Transactions on Consumer Electronics"}
@STRING{IEEE_J_IE         = "{IEEE} Transactions on Industrial Electronics"}
@STRING{IEEE_J_IECI       = "{IEEE} Transactions on Industrial Electronics and Control Instrumentation"}
@STRING{IEEE_J_IA         = "{IEEE} Transactions on Industry Applications"}
@STRING{IEEE_J_IGA        = "{IEEE} Transactions on Industry and General Applications"}



@STRING{IEEE_J_IM         = "{IEEE} Transactions on Instrumentation and Measurement"}



@STRING{IEEE_J_JEM        = "{IEEE/TMS} Journal of Electronic Materials"}
@STRING{IEEE_J_DEI        = "{IEEE} Transactions on Dielectrics and Electrical Insulation"}
@STRING{IEEE_J_EI         = "{IEEE} Transactions on Electrical Insulation"}



@STRING{IEEE_J_MECH       = "{IEEE/ASME} Transactions on Mechatronics"}
@STRING{IEEE_J_MEMS       = "{IEEE/ASME} Journal of Microelectromechanical Systems"}



@STRING{IEEE_J_BME        = "{IEEE} Transactions on Biomedical Engineering"}
@STRING{IEEE_J_B-ME       = "{IEEE} Transactions on Bio-Medical Engineering"}
@STRING{IEEE_J_BMELC      = "{IEEE} Transactions on Bio-Medical Electronics"}
@STRING{IEEE_J_ITBM       = "{IEEE} Transactions on Information Technology in Biomedicine"}
@STRING{IEEE_J_ME         = "{IEEE} Transactions on Medical Electronics"}
@STRING{IEEE_J_MI         = "{IEEE} Transactions on Medical Imaging"}
@STRING{IEEE_J_MCTE       = "{IEEE} Transactions on Molecular, Cellular and Tissue Engineering"}
@STRING{IEEE_J_NB         = "{IEEE} Transactions on NanoBioscience"}
@STRING{IEEE_J_NSRE       = "{IEEE} Transactions on Neural Systems and Rehabilitation Engineering"}
@STRING{IEEE_J_RE         = "{IEEE} Transactions on Rehabilitation Engineering"}



@STRING{IEEE_J_PTL        = "{IEEE} Photonics Technology Letters"}
@STRING{IEEE_J_JLT        = "{IEEE/OSA} Journal of Lightwave Technology"}



@STRING{IEEE_J_EDL        = "{IEEE} Electron Device Letters"}
@STRING{IEEE_J_JQE        = "{IEEE} Journal of Quantum Electronics"}
@STRING{IEEE_J_JSTQE      = "{IEEE} Journal of Selected Topics in Quantum Electronics"}
@STRING{IEEE_J_ED         = "{IEEE} Transactions on Electron Devices"}
@STRING{IEEE_J_NANO       = "{IEEE} Transactions on Nanotechnology"}
@STRING{IEEE_J_NS         = "{IEEE} Transactions on Nuclear Science"}
@STRING{IEEE_J_PS         = "{IEEE} Transactions on Plasma Science"}



@STRING{IEEE_J_DMR        = "{IEEE} Transactions on Device and Materials Reliability"}
@STRING{IEEE_J_R          = "{IEEE} Transactions on Reliability"}



@STRING{IEEE_J_ESSL       = "{IEEE/ECS} Electrochemical and Solid-State Letters"}
@STRING{IEEE_J_JSSC       = "{IEEE} Journal of Solid-State Circuits"}
@STRING{IEEE_J_ASC        = "{IEEE} Transactions on Applied Superconductivity"}
@STRING{IEEE_J_SM         = "{IEEE} Transactions on Semiconductor Manufacturing"}



@STRING{IEEE_J_SENSOR     = "{IEEE} Sensors Journal"}



@STRING{IEEE_J_VLSI       = "{IEEE} Transactions on Very Large Scale Integration ({VLSI}) Systems"}







@STRING{IEEE_M_AES        = "{IEEE} Aerospace and Electronics Systems Magazine"}
@STRING{IEEE_M_HIST       = "{IEEE} Annals of the History of Computing"}
@STRING{IEEE_M_AP         = "{IEEE} Antennas and Propagation Magazine"}
@STRING{IEEE_M_ASSP       = "{IEEE} {ASSP} Magazine"}
@STRING{IEEE_M_CD         = "{IEEE} Circuits and Devices Magazine"}
@STRING{IEEE_M_CAS        = "{IEEE} Circuits and Systems Magazine"}
@STRING{IEEE_M_COM        = "{IEEE} Communications Magazine"}
@STRING{IEEE_M_COMSOC     = "{IEEE} Communications Society Magazine"}
@STRING{IEEE_M_CSE        = "{IEEE} Computing in Science and Engineering"}
@STRING{IEEE_M_CSEM       = "{IEEE} Computational Science and Engineering Magazine"}
@STRING{IEEE_M_C          = "{IEEE} Computer"}
@STRING{IEEE_M_CAP        = "{IEEE} Computer Applications in Power"}
@STRING{IEEE_M_CGA        = "{IEEE} Computer Graphics and Applications"}
@STRING{IEEE_M_CONC       = "{IEEE} Concurrency"}
@STRING{IEEE_M_CS         = "{IEEE} Control Systems Magazine"}
@STRING{IEEE_M_DTC        = "{IEEE} Design and Test of Computers"}
@STRING{IEEE_M_EI         = "{IEEE} Electrical Insulation Magazine"}
@STRING{IEEE_M_EMB        = "{IEEE} Engineering in Medicine and Biology Magazine"}
@STRING{IEEE_M_EMR        = "{IEEE} Engineering Management Review"}
@STRING{IEEE_M_EXP        = "{IEEE} Expert"}
@STRING{IEEE_M_IA         = "{IEEE} Industry Applications Magazine"}
@STRING{IEEE_M_IM         = "{IEEE} Instrumentation and Measurement Magazine"}
@STRING{IEEE_M_IS         = "{IEEE} Intelligent Systems"}
@STRING{IEEE_M_IC         = "{IEEE} Internet Computing"}
@STRING{IEEE_M_ITP        = "{IEEE} {IT} Professional"}
@STRING{IEEE_M_MICRO      = "{IEEE} Micro"}
@STRING{IEEE_M_MW         = "{IEEE} Microwave Magazine"}
@STRING{IEEE_M_MM         = "{IEEE} Multimedia"}
@STRING{IEEE_M_NET        = "{IEEE} Network"}
@STRING{IEEE_M_PCOM       = "{IEEE} Personal Communications Magazine"}
@STRING{IEEE_M_POT        = "{IEEE} Potentials"}
@STRING{IEEE_M_PE         = "{IEEE} Power and Energy Magazine"}
@STRING{IEEE_M_PER        = "{IEEE} Power Engineering Review"}
@STRING{IEEE_M_RA         = "{IEEE} Robotics and Automation Magazine"}
@STRING{IEEE_M_SP         = "{IEEE} Signal Processing Magazine"}
@STRING{IEEE_M_S          = "{IEEE} Software"}
@STRING{IEEE_M_SPECT      = "{IEEE} Spectrum"}
@STRING{IEEE_M_TS         = "{IEEE} Technology and Society Magazine"}
@STRING{IEEE_M_WC         = "{IEEE} Wireless Communications Magazine"}
@STRING{IEEE_M_TODAY      = "Today's Engineer"}





\end{document}